\documentclass[sn-nature]{sn-jnl}

\usepackage{graphicx}%
\usepackage{multirow}%
\usepackage{amsmath,amssymb,amsfonts}%
\usepackage{amsthm}%
\usepackage{mathrsfs}%
\usepackage[title]{appendix}%
\usepackage{xcolor}%
\usepackage{textcomp}%
\usepackage{manyfoot}%
\usepackage{booktabs}%
\usepackage{algorithm}%
\usepackage{algorithmicx}%
\usepackage{algpseudocode}%
\usepackage{listings}%
\usepackage{subcaption}
\usepackage{placeins}

\begin{document}

\title[Article Title]{Multi-view biomedical foundation models for molecule-target and property prediction}


\author[1]{\fnm{Parthasarathy} \sur{Suryanarayanan}}
\equalcont{These authors contributed equally to this work.}

\author[2,3]{\fnm{Yunguang} \sur{Qiu}}
\equalcont{These authors contributed equally to this work.}

\author[4]{\fnm{Shreyans} \sur{Sethi}}

\author[1]{\fnm{Diwakar} \sur{Mahajan}}

\author[1]{\fnm{Hongyang} \sur{Li}}

\author[2]{\fnm{Yuxin} \sur{Yang}}

\author[1]{\fnm{Elif} \sur{Eyigoz}}

\author[1]{\fnm{Aldo} \sur{Guzm\'an S\'aenz}}

\author[1]{\fnm{Daniel E.} \sur{Platt}}

\author[1]{\fnm{Timothy H.} \sur{Rumbell}}

\author[5]{\fnm{Kenney} \sur{Ng}}

\author[1]{\fnm{Sanjoy} \sur{Dey}}

\author[1]{\fnm{Myson} \sur{Burch}}

\author[5]{\fnm{Bum Chul} \sur{Kwon}}

\author[1]{\fnm{Pablo} \sur{Meyer}}

\author[2,3,6]{\fnm{Feixiong} \sur{Cheng}}

\author[1]{\fnm{Jianying} \sur{Hu}}

\author*[1]{\fnm{Joseph A.} \sur{Morrone}} \email{jamorron@us.ibm.com}

\affil[1]{\orgname{IBM TJ Watson Research Center}, \orgaddress{\street{1101 Kitchawan Rd}, \city{Yorktown Heights},  \state{NY}, \postcode{10598}, \country{USA}}}

\affil[2]{\orgdiv{Cleveland Clinic Genome Center}, \orgname{Lerner Research Institute, Cleveland Clinic}, \orgaddress{ \city{Cleveland},  \state{OH}, \postcode{44195}, \country{USA}}}

\affil[3]{\orgdiv{Genomic Medicine Institute}, \orgname{Lerner Research Institute, Cleveland Clinic}, \orgaddress{ \city{Cleveland}, \state{OH},  \postcode{44195}, \country{USA}}}

\affil[4]{\orgname{IBM Research - Almaden}, \orgaddress{\street{650 Harry Rd}, \city{San Jose},  \state{CA}, \postcode{95120}, \country{USA}}}

\affil[5]{\orgname{IBM Research}, \orgaddress{\street{314 Main St}}, \city{Cambridge},  \state{MA}, \postcode{02142}, \country{USA}}

\affil[6]{\orgdiv{Department of Molecular Medicine}, \orgname{Cleveland Clinic Lerner College of Medicine, Case Western Reserve University}, \city{Cleveland}, \state{OH}, \postcode{44195}, \country{USA}}


\abstract{Quality molecular representations are key to foundation model development in bio-medical research. Previous efforts have typically focused on a single representation or molecular view, which may have strengths or weaknesses on a given task.  We develop Multi-view Molecular Embedding with Late Fusion (MMELON), an approach that integrates graph, image and text views in a foundation model setting and may be readily extended to additional representations.  Single-view foundation models are each pre-trained on a dataset of up to 200M molecules.  The multi-view model performs robustly, matching the performance of the highest-ranked single-view.  It is validated on over 120 tasks, including molecular solubility,  ADME properties, and activity against G Protein-Coupled receptors (GPCRs). We identify 33 GPCRs that are related to \emph{Alzheimer's disease} and employ the multi-view model to select strong binders from a compound screen. Predictions are validated through structure-based modeling and identification of key binding motifs.}




\maketitle

\section{Introduction}\label{sec1} 

Drug discovery is a complex, multi-stage process.   Lead identification and lead optimization remain costly  with low success-rates~\cite{hughes_principles_2011}.  The prediction of a broad range of chemical and biological properties of candidate molecules is an essential component of screening and assessing molecules and data-driven, machine learning approaches have long aided in this process~\cite{zeng_deep_2022,cheng_artificial_2024,butkiewicz_benchmarking_2013, vamathevan_applications_2019, muratov_qsar_2020}.

Molecular representations form the basis of machine learning models~\cite{wigh_review_2022,zeng_deep_2022}.  However, learning useful and generalized latent representation is a hard problem due to limited amounts of labeled data, wide ranges of downstream tasks, vast chemical space, and large heterogeneity in molecular structures. Learning latent representations using unsupervised techniques is vital for such models to scale. Large language models (LLMs) have revolutionized many fields~\cite{zhao2024surveylargelanguagemodels} and similar sequence-based foundation models have shown promise to learn molecular representations and be trainable on many downstream property prediction tasks~\cite{chithrananda_chemberta_2020,chilingaryan_bartsmiles_2022,ross_large-scale_2022}.  A key advantage is that the transformer based architecture can learn in a self-supervised fashion to create a ``pre-trained'' molecular representation.   The most direct application of LLM like transformers is facilitated by a sequence, text-based representation (e.g. SMILES).

While sequence-based representations have been successful in capturing human-generated knowledge/information in LLMs, when it comes to representing natural objects like molecules, it is likely that sequence molecular representations alone will not be sufficient. Here we propose \textbf{M}ulti-view \textbf{M}olecular \textbf{E}mbedding with \textbf{L}ate Fusi\textbf{on} (MMELON) a flexible approach to aggregate multiple views in a foundation model setting and show that the resulting richer representations lead to robust performance across a large range of downstream tasks. The molecule use case demonstrates the effectiveness of the  approach, which can be easily extended to include proteins and other biological entities.  Our multi-view model is available on GitHub (\url{https://github.com/BiomedSciAI/biomed-multi-view}) and Hugging Face (\url{https://Huggingface.co/ibm-research/biomed.sm.mv-te-84m}). 

For molecules, the reason sequence alone may fall short is that intuitively, molecular representations should reflect the intrinsic geometry of the entities, including symmetries and similarity relations. Such geometry and composition give rise to chemical and biological properties~\cite{atz_geometric_2021}.  Indeed, traditional cheminformatics  makes use of molecular fingerprints~\cite{carhart_atom_1985,durant_reoptimization_2002, rogers_extended-connectivity_2010, oboyle_comparing_2016, rdkit}, akin to hand-engineered features in other fields.

Graph representations are naturally conducive to capturing molecular symmetries and other structural properties in a flexible manner~\cite{NIPS2015_f9be311e,altae-tran_low_2017,morrone_combining_2020,wieder_compact_2020,zeng_deep_2022}.  However  graph neural network (GNN) architectures lack the capacity to scale to large numbers of parameters and learn from very large datasets. Architectures that combine elements of graphs and transformer based approaches are better candidates for foundation models.  Graph-transformers can represent graphs in a high-capacity neural network and several variants of this approach have already been explored~\cite{kim_pure_2022,muller_attending_2023}.

Image-recognition is another field where molecular representations can draw from.  Image-based representation paired with a convolutional neural network (CNN) architecture have  been explored with promising results~\cite{goh2017chemceptiondeepneuralnetwork}.  ImageMol, an exemplar of this approach, shows strong performance across many downstream tasks~\cite{zeng_accurate_2022}. 

Each representation/modality described above may be considered a ‘view’ of the same molecule which can provide complementary information. A promising approach to achieve more comprehensive molecular representations is identifying effective ways of combining these views through multi-modal fusion. Multi-modal fusion is an active area of machine learning across many disciplines. While the molecular use-case has been explored in several publications, primarily in the context of combining graph with a text-based model or 2-dimensional and 3-dimensional molecular representations~\cite{liu_pre-training_2022,zhu_unified_2022,zhu_improving_2022,du_fusing_2023}, the scaling to large pre-training datasets ($>100 \text{M}$) and the quantification of the value of such combinations on downstream drug discovery tasks remain open to date.  

Our multi-view molecular foundation model architecture, MMELON (Fig. \ref{fig:modelembeds}A), leverages the suite of Biomedical Foundation Model (BMFM) technologies~\cite{bmfm_blog}.  This multi-view architecture combines molecular representations to learn a joint base multi-modal embedding. This base embedding is then used to fine tune downstream tasks for chemical and biological property prediction.  The model currently integrates three views: image, 2-dimensional chemically-bonded graph and text, although our architecture is extensible and can easily include other modalities such as 3-D conformations.  For the image view we adopt ImageMol~\cite{zeng_accurate_2022}, a state of the art molecular foundation model trained on images of molecules. Graph and text views are each pre-trained on a 200M molecular dataset curated from public sources~\cite{kim_pubchem_2022, tingle_zinc-22free_2023}. Using a pre-trained aggregator module to combine the embeddings, we show that the individual modalities contain complementary information.
The weight of each view is a learned parameter, yielding a transparent interpretation. Our multi-view foundation model is fine-tuned on an extensive set of downstream tasks for molecular property prediction~\cite{cheng_classification_2011,MoleculeNet,davis_comprehensive_2011}.  

The multi-view model is further validated by applying it to identity binders to  Alzheimer's disease (AD)-related targets. AD is a progressive age-related disorder that is anticipated to affect over 150 million people by 2050~\cite{2024AD}.  Current AD drugs remain limited whereas two antibody-based drugs that are recently approved by the FDA primarily target the early-stage of AD~\cite{Cummings2024Alzheimer}.  This challenge is largely due to the lack of potential druggable AD targets beyond the well-established targets, such as amyloid-beta (A$\beta$) or apolipoprotein E (APOE)~\cite{Qiu2024Artificial}. Emerging machine learning algorithms have greatly empowered target identification and small molecule discovery~\cite{cheng_artificial_2024}. For example, NETTAG~\cite{Xu2022Interpretable} provided key insights for treating AD by unraveling potential therapeutical targets, including G Protein-Coupled receptors (GPCR)~\cite{Thathiah2011role}. LISA-CPI~\cite{yang2024deep} identified promising small molecule treatments for pain by leveraging both target structure and ligand image information. Additionally, MRL-Mol~\cite{yang2024deepm} offered a preliminary investigation on the enhancement of deep learning models by integrating different molecular data modalities. Our multi-view model is fine-tuned on binding assay data collected for over 100 GPCRs~\cite{chembl2011,glass2015}. GPCRs associated to AD are identified and employed to discover strong binders from a set of gut metabolites and FDA approved drugs~\cite{drugbank2018}.  The predictions are validated by structure-based \textit{in silico} experiments that identify pharmacophores and binding modes.

\section{Results}\label{sec2}
\subsection{Overview of models and architectures}\label{pret}
We adopt three single view architectures and models based on image, text and graph molecular representations. The models are herein referred to as Image, Graph, and Text, respectively. A schematic for the pre-training strategies we employ is shown in Fig. \ref{fig:modelembeds}B.  For Image, we adopt the ImageMol model~\cite{zeng_accurate_2022}.  ImageMol utilizes a CNN-based architecture and is pre-trained on $10 \text{M}$ compounds selected from PubChem~\cite{kim_pubchem_2022}. The pre-trained model and weights are taken from ImageMol's publicly released checkpoint.  This checkpoint is fine-tuned in Sec. \ref{baseline_results} using our independently developed code that is utilized here. For Text, our architecture and pre-training strategy is adopted from MolFormer~\cite{ross_large-scale_2022}.

The Graph architecture is adopted from TokenGT~\cite{kim_pure_2022}. This architecture processes chemically-bonded graphs as a sequence of tokens before input to the transformer.  We design three pre-training tasks to pre-train the graph-transformer: node feature masking, edge prediction, and Betti number prediction. The last task predicts graph-topological features and is, to our knowledge, novel in this field.  More details of Graph pre-training are given in Sec. \ref{methods_graph}.

We pre-train the Graph and Text on a set of 200M molecules curated from two public data sources: PubChem and ZINC22~\cite{kim_pubchem_2022,tingle_zinc-22free_2023}.  The PubChem dataset contains a broad distribution of molecules including a minor population that is rather large and flexible, and as such of less interest for small molecule drug discovery.  We filter out this minor population using a drug likeliness criteria~\cite{LIPINSKI20013}. An additional $120 \text{M}$ molecules are randomly sampled from ZINC22.  This set of molecules covers a distribution of drug like molecules and molecular properties (see Sec. \ref{methods_pretrain_data} and Figs. S1-S2).  Graph and Text are pre-trained for $3$ epochs each.

There are a number of multi-modal fusion strategies that may be employed to combine representations~\cite{liang_foundations_2023,liu_pre-training_2022,zhu_unified_2022,du_fusing_2023}.  We adopt a late fusion approach because it readily integrates separately pre-trained models and naturally supports analyses to gain insights into the role and importance of different modalities for different downstream tasks. In this approach, the singe-views are first separately trained and then combined using an aggregator sub-network (see Fig. \ref{fig:modelembeds}A).  The single-view foundation models serve as pre-trained encoders that are utilized for both aggregator pre-training (see Sec. \ref{methods_agg}) and downstream task fine-tuning (see Sec. \ref{baseline_results}). For our aggregator architecture, we use an attention-based approach inspired by Ref. \cite{zhu_improving_2022} where the contribution of the $m^\text{th}$ view to the overall embedding vector, $z_i^\text{mv}$, is weighted by coefficient $\alpha^m$, 
\begin{align}
z_i^\text{mv} &\propto \sum\limits_{m\in\mathcal{M}} \alpha^m z^m_i  \label{mmf_eqn_main}
\end{align}
One advantage of the MMELON approach, is that the weight $\alpha$ is directly related to the relative importance of each view in contributing to downstream tasks. The performance of other alternative approaches to our chosen aggregator on fine-tune tasks were explored.  A more detailed discussion of the aggregator sub-network is given in Sec. \ref{methods_agg}.

\subsection{Pre-trained models learn comprehensive embeddings} \label{pretrain_results}
\begin{figure}
\centering
\includegraphics[width=0.8\textwidth]{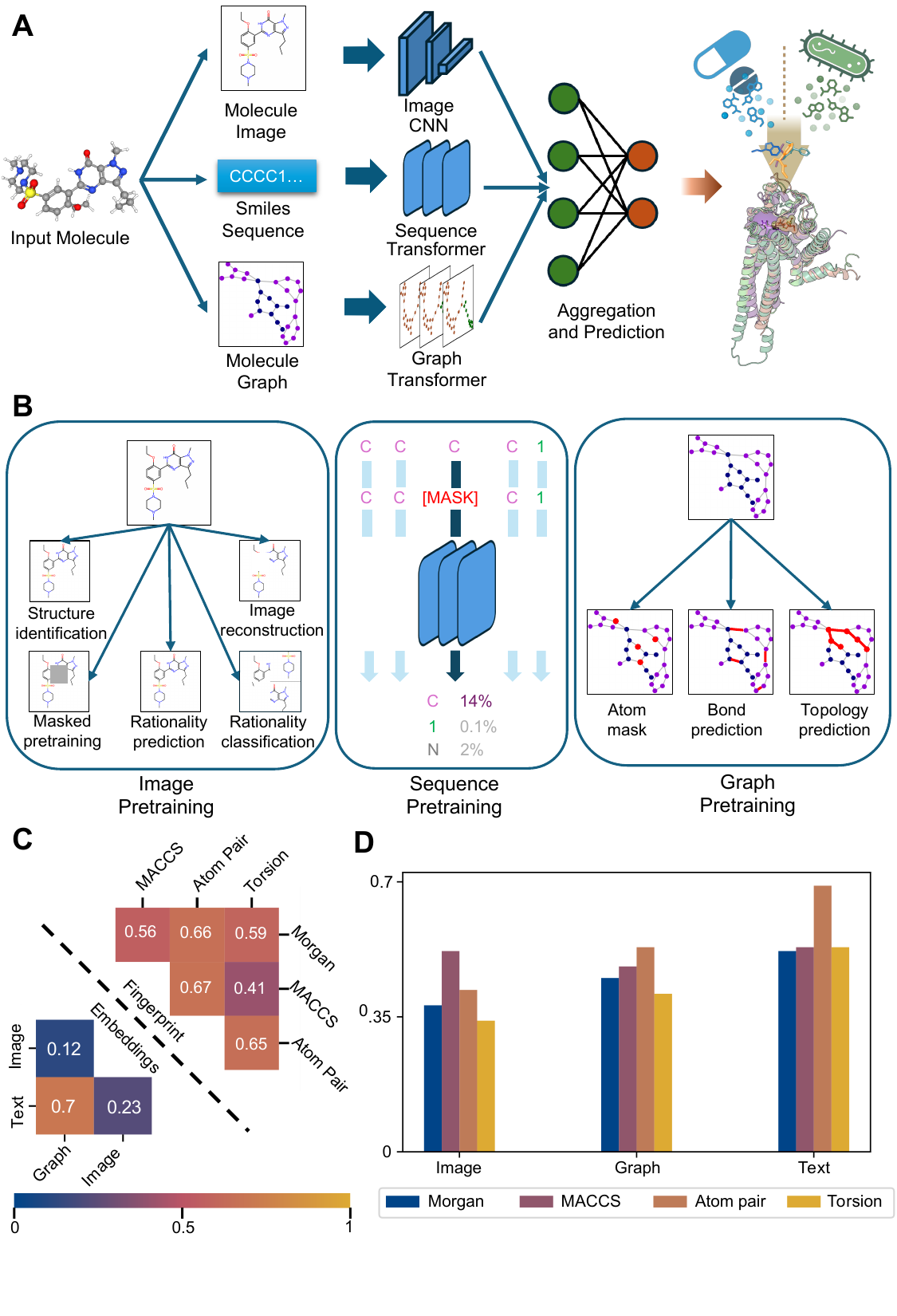}
\caption{(A) Schematic of multi-view architecture.  Embeddings from three single view pre-trained encoders (Image, Graph and Text) are combined by the aggregator module into a combined embedding. The network is fine-tunable for downstream predictive tasks. (B) Pre-training strategies for single view (Image, Text (sequence) and Graph) embeddings.   These strategies are described in detail in Sec. \ref{methods_graph}.  (C) The elements of the correlation matrix between the three-single view embeddings (lower triangle) and between four fingerprints (upper triangle). (D)  The correlation between four fingerprints and the pre-trained, Image (left grouping), Graph (center grouping) and Text (right grouping) embeddings.\label{fig:modelembeds}}
\end{figure}

Image, Graph and Text offer intrinsically different views of chemical structure.  However, it is expected that there is at least some overlap in their description of the embedding space.  We sample a subset of $10,0000$ molecules from the pre-training set to perform the following analysis.  In order to characterize the pre-trained embeddings, we compute the correlation between their Euclidean distances in Fig. \ref{fig:modelembeds}C.    It can be seen that of the three single-view embeddings, Image is the most distinct.  Text and Graph are highly correlated to each other ($c=0.7$), indicating that these two representations have a comparable description of molecular similarity.

Pre-trained molecular embeddings should ideally capture important aspects of the chemical structures of the underlying molecules. While there is no absolute way of measuring this, comparison with fingerprints that have been developed by domain experts can serve as a sanity check on whether pre-trained embeddings are in line with chemical knowledge ~\cite{ross_large-scale_2022}.  Four common fingerprint descriptors (Morgan, atom pair, MACCS and topological torsion) are chosen for this purpose~\cite{nilakantan_topological_1987,carhart_atom_1985,durant_reoptimization_2002,rogers_extended-connectivity_2010}.  We first compute the correlation between fingerprints, and while the atom pair fingerprint is most correlated with the others, overall the chosen descriptors are moderately correlated and yield different descriptions of the molecules (see Fig. \ref{fig:modelembeds}C).

The correlations are computed between the Euclidean distance of the pre-trained models and Tanimoto distance of the fingerprints and plotted in Fig. \ref{fig:modelembeds}D.  It can be seen that there is moderate correlation between pre-trained embedding and the fingerprints, demonstrating the validity of the embeddings. Overall, Text exhibits the highest correlation, followed by Graph and Image. Image is most correlated to the MACCS fingerprint, which is expected given that it is used in one of the ImageMol model's pre-training tasks~\cite{zeng_accurate_2022}.  Finally, we note that the correlation with fingerprints only serves as a sanity check and strong agreement with fingerprints is not necessarily desirable for all down stream property prediction tasks, because each fingerprint is tailored using a limited set of hand-engineered features. A key objective of learning pre-trained multi-view embeddings is to achieve richer, more flexible representations.

\FloatBarrier

\subsection{Baseline performance: fine-tuning on downstream tasks }\label{baseline_results}
The prediction of electronic properties, ligand-protein binding affinity, pharmokinetic properties, and toxicity are examples of tasks that are essential to chemistry and drug discovery.  MoleculeNet~\cite{MoleculeNet} is a collection of datasets covering such tasks that serves as a common test for machine learning models. Performance on this suite of benchmarks is neither decisive, nor necessarily reflective of all important use-cases in computational drug discovery, but currently it stands as one of the most widely used point of comparison between models.  We also assess our model on predicting the inhibition of cytochrome P-450 (CYP) isoforms~\cite{cheng_classification_2011} which are critical in drug metabolism pathways, and the recently released ComputationalADME dataset, which include tasks such as liver microsomal stability and plasma protein binding~\cite{computationalADME2023}.  In addition to the choice of dataset, the choice of data splitting is critical in robust assessment of model performance.   We primarily adopt scaffold-based methodologies, but also utilize random splits where appropriate (see SI for further discussion).

\begin{figure}
    \centering
    \includegraphics[width=1\linewidth]{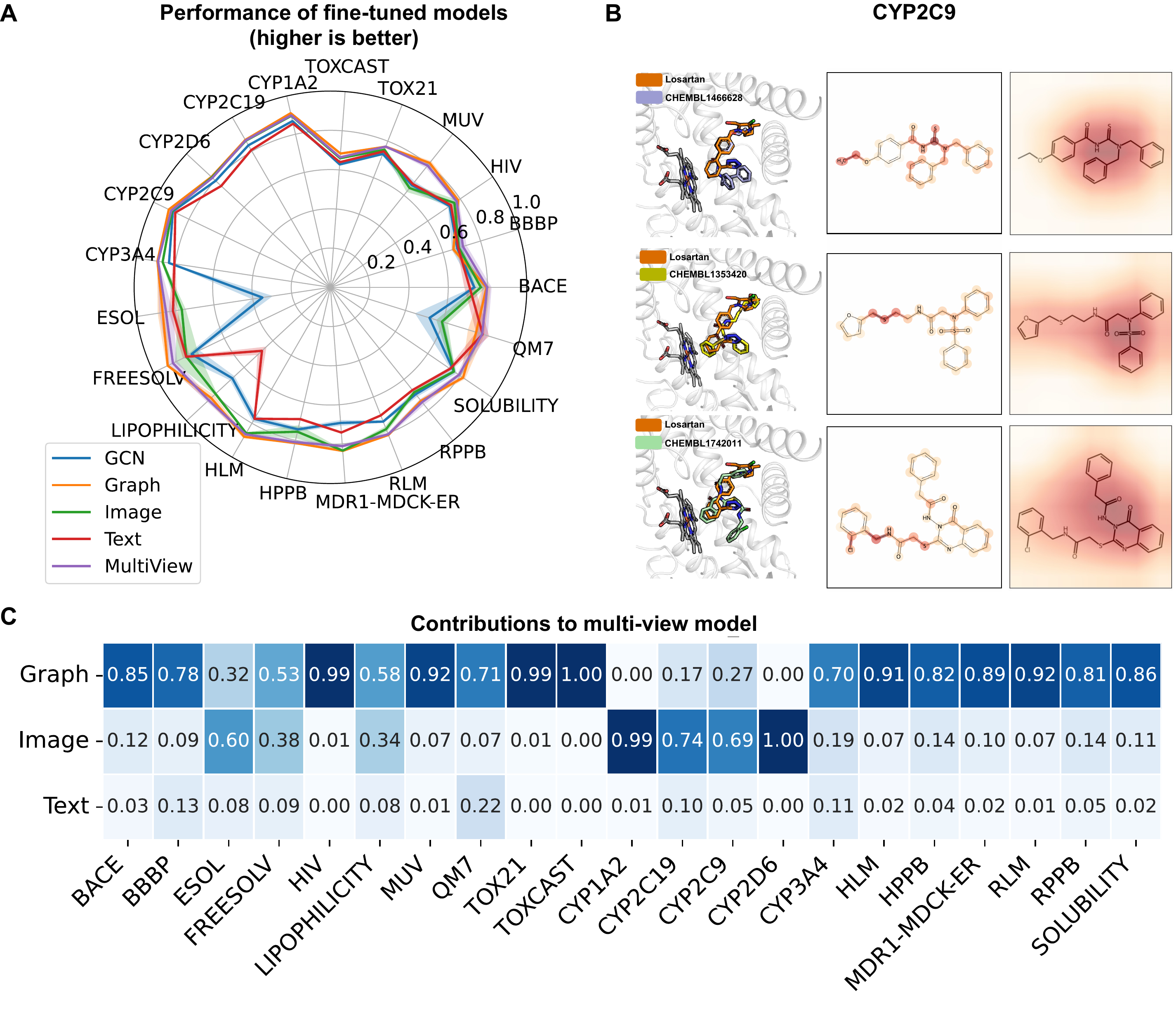}
    \caption{(A) Performance of GCN (blue line), Graph (orange line), Image (green line), Text (red line) and multi-view model (purple line) fine-tuned on diverse downstream tasks.  Train, validation, and test splitting is performed as described in the text.  Classification tasks are characterized by ROC-AUC.  Regression metrics (RMSE, MAE) are scaled so that they can be plotted with ROC-AUC values.  (B) Docking poses of selected bioactive molecules against CYP2C9 as predicted by the multi-view model are compared with the reported crystal structure (PDB: 8VX0, left column panels) and are shown alongside graph (center column panels) and image (right column panels) attention level heatmaps. Darker color indicates higher attention level.  The heme group and remaining catalytic cite are rendered using the licorice and the ribbon style, respectively. (C) Heat map of the weights $\alpha$ that the multi-view model assigns to Graph, Text and Image model for a subset of tasks shown in panel A.}
    \label{fig:molnet_scaffold}
\end{figure}

In Fig. \ref{fig:molnet_scaffold}A, we show the performance of our multi-view model on 21 tasks from MoleculeNet, CYP and ComputationalADME.  In order to evaluate the contributions from each view/modality, we focus on comparisons of single view Graph, Text, and Image models and the multi-view model.  To serve as a simple baseline, results from a graph convolutional network (GCN) implemented with PyTorch Geometric are shown~\cite{fey2019fastgraphrepresentationlearning}.  All are fine-tuned in the same code base with hyper-parameter optimization (primarily the learning rate, see Sec. \ref{methods_ft}). The area under the receiver-operator curve (ROC-AUC) is the chosen metric for the classification tasks and root mean square error (RMSE) or mean absolute error (MAE) is chosen for regression~\cite{MoleculeNet}.  Regression metrics are scaled in Fig. \ref{fig:molnet_scaffold}A so as to be plotted alongside ROC-AUCs (see Sec. \ref{methods_ft}). Tables S2-S5 report the numerical values for experiments with $95\%$ confidence intervals.   A detailed description of each task is given in Table S1.

We find that overall that the Graph model outperforms Image and Text on our benchmark tests.  The Graph model performs marginally worse than other single-views in only a handful of examples (e.g., BBBP, QM7).  We find that the multi-view model performs robustly across all tasks, spanning areas such as molecule solubility prediction (ESOL, Lipophilicity), binding activity assay-based classification (HIV, MUV) and ADME-Tox based tasks.  By `robustness' we indicate that the multi-view mode yields similar performance to the highest-ranked single view and does not produce a poor result on any tested dataset (see Fig. \ref{fig:molnet_scaffold}A). 

One advantage of the multi-view late-fusion model architecture is that the contributions of single views to each task can be readily recovered through the aggregator weights, $\alpha$ (see Eqn. \ref{alpha_eqn}).  The $\alpha$ values are visualized as a heat map for MoleculeNet tasks in results in Fig. \ref{fig:molnet_scaffold}C.   We note that the pre-trained aggregator (see Sec. \ref{methods_agg}) is free to evolve during fine-tuning and adapt to the downstream tasks.  The Graph model is heavily weighted for most tasks, which is consistent with the observation that Multi-View performance is generally similar to Graph.  A complementary analysis measuring the agreement between different views of selected fine-tuned tasks is given in Sec. \ref{methods_statistics} and Figs. S4-S5.  For CYP models, both Graph and Image are nearly equally correlated with the multi-view model, and this degeneracy can be related to trends in Fig. \ref{fig:molnet_scaffold}C.

With the exception of QM7, the Text Model is a minor contributor to the multi-view fine-tuned embeddings and while the predictions between text and multi-view are correlated, they exhibits the weakest relation (see Figs. S4-S5).  This could be due to it being similar to Graph (see Sec. \ref{pretrain_results}) and yet the Graph model is a stronger single view model. This observation requires further study and could be related to previous results shown for multi-modal architectures~\cite{zhu_improving_2022}.  Overall, it can be stated that Graph and Image represent more distinct embeddings (see Sec. \ref{pretrain_results}) and are more complementary.  

One set of tasks we highlight is cytochrome-P450 (CYP) inhibition for five CYP isoforms~\cite{cheng_classification_2011}. Performance as measured by ROC-AUC is high, ranging from $0.90$ to $0.82$ and results are comparable with SOTA performance~\cite{zeng_accurate_2022}. For example, isoform CYP2C9 is one of most important contributor to drug metabolism~\cite{Zhao2021Cytochrome} and our model achieved a high ROC-AUC value (0.90, Fig. \ref{fig:molnet_scaffold}A). The binding modes of top bio-active molecules computed from molecular docking align well with reported structures of CYP2C9 in complex with the drug losartan~\cite{PARIKH2024112622} (PDB ID: 8VX0, see Fig. \ref{fig:molnet_scaffold}B).  In addition, the salient molecular moieties as predicted by the graph and image views of multi-view are both close to the catalytic site of CYP2C9~\cite{Parikh2024Structural}, suggesting our multi-view models could highlight the important binding features.

In addition, we consider a task that combines our multi-view embedding of small molecules with a popular sequence-based protein embedding, ESM~\cite{Rao2020.12.15.422761}.  It is an open question what the protein embedding contributes to so-called `drug-target interaction' tasks~\cite{sieg2019need,morrone_combining_2020, gorantla_proteins_2024}.  However, as ligand embeddings are known to drive model performance, this could still serve as another test of our small-molecule models.  We fine-tuned our multi-view and the single-view models on the Davis dataset~\cite{davis_comprehensive_2011} describing the binding of 72 kinase inhibitors with 442 kinases.  We find that the performance of our model exceeds that of a SOTA GNN implementation~\cite{gorantla_proteins_2024} (Fig. S3).

\FloatBarrier

\subsection{Case study: GPCR metabolite and drug target binding and identification}\label{case_results}
While benchmarks may validate models, it is key to demonstrate the use of molecular foundation models in a wider context.  Here we use our multi-view model to conduct a large-scale virtual screening experiment by fine-tuning small molecule binding affinity models for 106 G Protein-Coupled receptors (GPCRs).  We choose the multi-view model based its potential for robust performance over a large number of unexplored tasks.

GPCRs are important group of targets for drug discovery campaigns~\cite{hauser2017trends}. A recent study by some of us identified a selection of GPCRs as potential drug targets for Alzheimer's disease~\cite{qiu_systematic_2024}.  We consider the intersection between our screening database and AD-related GPCRs to uncover potential binders.

\begin{figure}[h]
    \centering
    \includegraphics[width=1.0\linewidth]{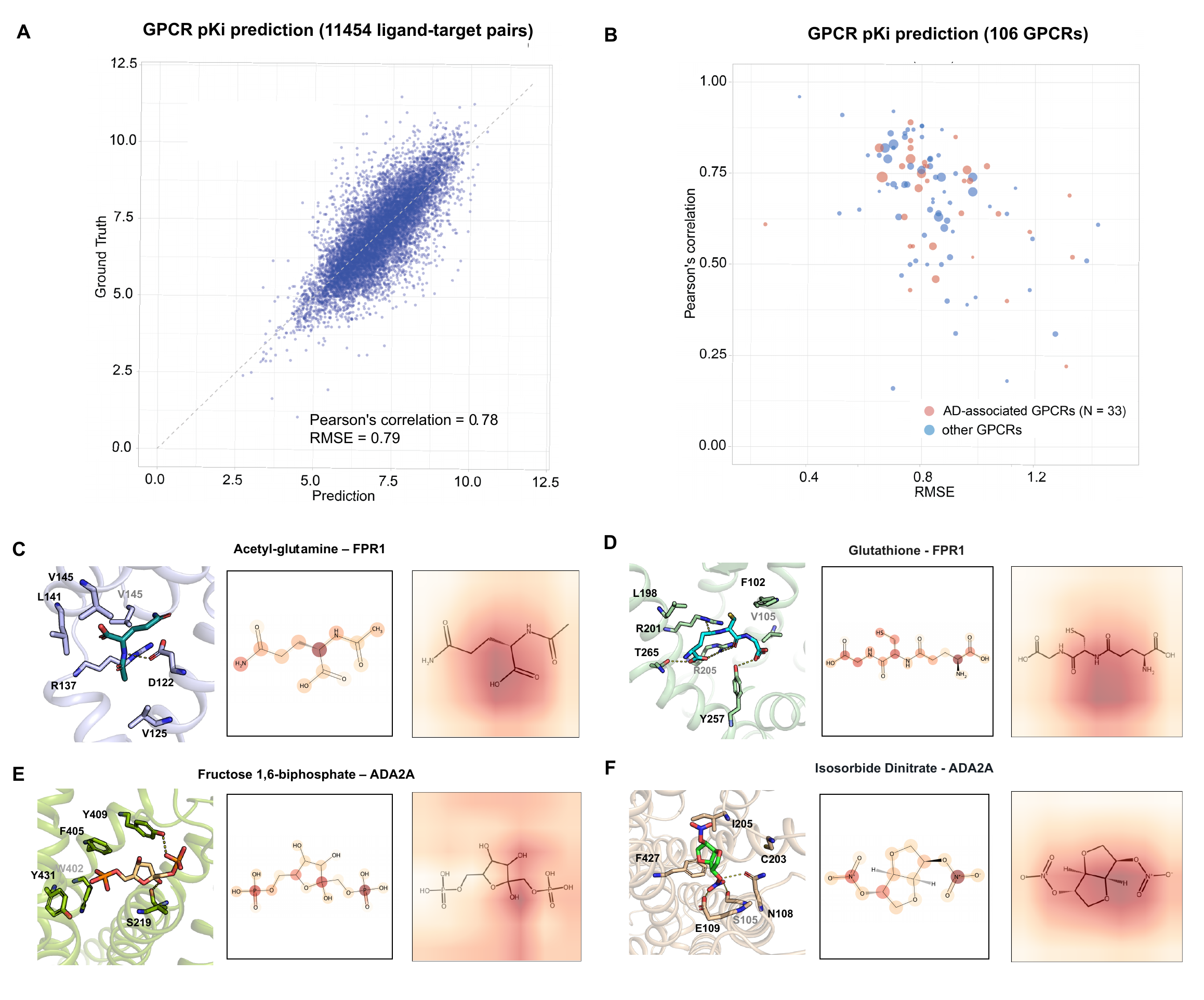}
    \caption{ (A) Correlation plot of predicted vs. experimental pKI values in the held out set.  Values of the Pearson correlation and the RMSE are given in the legend.  (B) Pearson correlation coefficient plotted against RMSE for 33 Alzheimer's disease related GPCR targets (red dots) and others (blue dots).  Examples of strong binders to FPR1 (C-D) and ADA2A (E-F) are depicted in the respective panels.  Binding poses generated from docking software (left column panels) are depicted alongside heat maps on attention levels as predicted by the graph (center column panels) and image (right column panels) sub-networks of our model. Darker color indicates higher attention level.}
    \label{fig:gpcr}
\end{figure}

We fine-tune our multi-view model on experimental binding affinity assays for 106 GPCR targets.  This dataset contains experimental pKI values curated from ChEMBL~\cite{chembl2011} and GLASS~\cite{glass2015}. All GPCRs datasets were retrieved but the GPCRs with fewer bioactivity measurements ($< 100$) were not considered. Before training, the dataset is filtered to exclude molecules with a molecular weight $>600$ Da.  This cutoff is commiserate with the molecular properties of the metabolites and drug molecules that are in the screening (inference) set.  Datasets with less than 50 data points were removed from the set.  Each assay/GPCR target is treated as a separate regression task and data is split using the balanced scaffold protocol~\cite{zeng_accurate_2022}.    In Fig. \ref{fig:gpcr}A, we plot the prediction on the held-out set against the experimental values.  The multi-view fine-tuned models are overall in good agreement with the held-out set (Pearson correlation coefficient 0.78, average RMSE 0.79).  To further validate our model, in Fig. \ref{fig:gpcr}B  we plot the Pearson correlation coefficient against the RMSE for each individual target. These metrics are negatively correlated, as expected.  Over $80\%$ of fine-tuned models have $\text{RMSE} < 1$ and nearly $70\%$ have correlations $> 0.6$  on the held-out set.  

To unravel AD-related GPCRs, genetics-informed causal proteins inferred by Mendelian Randomization (MR) based on Genome-Wide Association Studies (GWAS) \cite{Jansen2019Genome, Wightman2021genome} and multi-omics (transcriptomics and proteomics)-informed differential expressed genes (DEGs) on the human brain~\cite{Zhou2021AlzGPS} were inspected. Strong multi-omics evidence-supported protein was defined as those that were differential expressed in as least 5 datasets we previously compiled~\cite{Zhou2021AlzGPS} (see Fig. S6 and Table S6). Next we sought to identify potential safe AD therapies from gut metabolites and FDA-approved drugs. The validated multi-view model is used to predict the binding affinity of a set of 515 gut metabolites~\cite{Han2021metabolomics} and 2,504 FDA-approved drugs from DrugBank (Version 2021.1)~\cite{drugbank2018} The red dots in Fig. \ref{fig:gpcr}B indicate the performance on the 33 AD-related GPCRs is within range of the larger set (Pearson correlation coefficient: $ 0.67 $, average $\text{RMSE}: 0.88)$. 

Next we consider interactions of molecules in our virtual screen with specific AD-related receptor. FPR1 is suggested as a genetics-informed GPCR. High level of FPR1 is causally associated with increased risks of AD (Beta = -0.1, FDR  = 0.001, Fig. S6A). Through inspecting top 10 small molecules in our screen, we found that a gut metabolite, acetyl-glutamine is predicted to interact with FPR1 (multi-view score: 6.8). By re-analyzing the abundance of gut metabolites in microbial strains \textit{in vitro}~\cite{Han2021metabolomics}, acetyl-glutamine was determined to have a high level in \textit{Ruminococcus gnavus} (log2FoldChange: 1.19), a strain that mono-colonized in mice performed better on a spatial working memory test~\cite{Coletto2022role}. Additionally, over-expression of $\text{A}\beta$ protein is significantly characterized by the decreased \textit{Ruminococcus gnavus}~\cite{dos2020Impact}. Three-dimensional (3D) binding interaction analysis revealed that acetyl-glutamine is located in an allosteric binding site that is distinct from the classical binding site, indicating a potential regulation mechanism (Fig. \ref{fig:gpcr}C).  For top predicted drugs, glutathione (GSH), an anti-oxidant drug \cite{Averill2023antioxidant}, GSH, is prioritized (multi-view score: 6.53) to interact with FPR1. GSH is a nutrition supplementation that consist of three amino acids, including glutamic acid \cite{Gould2019Impact}. The depletion of GSH has been found in the hippocampal regions in mild cognitive impairment (MCI) and AD patients compared to healthy control~\cite{Mandal2019Cognitive}. GSH forms strong hydrogen bond interactions with FPR1 in the classic binding site (Fig. \ref{fig:gpcr}D).   We find that image ($\alpha=0.51$) and graph ($\alpha=0.39$) are the top contributors to the FPR1 multi-view model and generate attention heat maps from the image and graph sub-networks (Fig. \ref{fig:gpcr}C-D). Most of the functional groups of acetyl-glutamine (\textit{e.g.}, amide group) and GSH have high attentions.  For acetyl-glutamine binding to FPR1, the image highlights the attention of amide group that forms hydrogen bond with FPR1 while the graph captures the main chain carbon that controls the direction of amide group, which is consistent with the binding mode.

ADA2A is a GPCR with the strong multi-omics evidence to have a significant lower expression in AD (differential expressed in 1 human and 4 mouse transcriptomics datasets, Fig. S6B)~\cite{Zhou2021AlzGPS}. A gut metabolite, fructose 1,6-biphosphate, is predicted to strongly interact with ADA2A (multi-view score 7.37). Fructose 1,6-biphosphate is involved in glycolysis, the level of which was down regulated in APP/PS1 mice~\cite{Minhas2024Restoring}. High levels of 1,6-biphosphate have been detected in genus \textit{Mitsuokella} (log2FoldChange: 13.42), which is decreased in $\text{A}\beta+$ (positive) patients~\cite{Jung2022Gut}. Fructose 1,6-biphosphate is predicted to bind a classic site of ADA2A (Fig. \ref{fig:gpcr}E). Isosorbide dinitrate, a drug used to prevent and treat chest pain caused by coronary artery disease\cite{drugbank2018}, is predicted as a top candidate to bind ADA2A (multi-view score: 8.21, Fig. \ref{fig:gpcr}F). The high attention of fructose 1,6-biphosphate and isosorbide dinitrate were captured for key chemical moieties (Fig. \ref{fig:gpcr}E-F). Although the image serves as the primary contributor ($\alpha=0.76$), the graph representations play a crucial supplementary role ($\alpha=0.22$) to preserve functional atoms, such as phosphorus atoms in phosphate of fructose 1,6-biphosphate and nitrogen atoms in nitro groups of isosorbide dinitrate, all of which establish key interactions with ADA2A. 

\FloatBarrier

\section{Discussion}\label{conclusion}
A molecular foundation model that is fine-tunable on a broad, diverse set of tasks can serve as a crucial component of any discovery workflow and the choice of molecular representation is crucial.  Each choice may have strengths and weaknesses, and each individual representation (view)  can perform well on certain downstream tasks and not others. Here we introduce MMELON, a multi-view architecture that combines molecular representations from multiple embeddings using an interpretable late fusion approach, utilizing foundation models based on three common representations of chemical structure, Graph, Image and Text.

We fine-tune our multi-view model on a set of benchmark datasets that span diverse tasks including binding activity, drug metabolism, and ADME properties.  We compare the performance of the multi-view model with the single view models whose encodings feed into it.  Overall the multi-view model performs robustly across a wide range of property prediction tasks involving both classification and regression, performing best or next best on any nearly every considered task and closely matching the best single-view result.  

We note that the late fusion approaches considered here do not clearly surpass the strongest single view on the set of chosen benchmarks.  Indeed, the weights of each view extracted from our model indicate the fine-tuned multi-view models are generally dominated by a single contribution (Fig. \ref{fig:molnet_scaffold}C).  Although the embedding analysis in Sec. \ref{pretrain_results} suggests complementarity (or at minimum weak correlation) between the Image and Graph pre-trained representations, we find that the fine-tuning process introduces additional correlations between task-specific models (see Figs. S4-S5).  Overfitting of chemistry models to a select modality or property has generally been attributed to the choice of  dataset~\cite{sieg2019need,chen2019hidden,morrone_combining_2020,gorantla_proteins_2024}, yet behavior similar to that observed here have been reported in other fields of machine learning and said to arise from imbalanced learning between modes~\cite{wang2020makes,wu2022characterizing,zhang2024multimodal}.  Here, we experiment with multi-learning rate strategies (see Methods) and future work will further explore alternative training and fusion strategies.

In our benchmark results, we find the multi-view model largely mirrors the Graph model performance, which is the highest-ranked single-view model across the studied tasks.  In our hands, it yields stronger performance overall than our Text and Image models.  The Graph model is considered to be a more natural representation of the symmetries of the molecular graph and better capture the sub-structures that drive recognition~\cite{morrone_combining_2020}.  For the Graph model we introduce a novel pre-training task, Betti number prediction that captures topological features that complement the other, highly localized edge and node-based pre-training tasks.  However, while selected  a large and diverse set of benchmark tests, no experiment is fully exhaustive and these observations are not guaranteed to be valid across all potential tasks.  The multi-view approach is therefore useful to ensure against such variability in cases where a single view is not superior.

To further validate the effectiveness of MMELON we apply our pre-trained multi-view model to a large collection of GPCR assays to create over $100$ fine-tuned activity models. This serves as a virtual screening panel against which a set of metabolites and FDA approved drugs are screened.  We further zoom in to a large number (33) of GPCR targets that are implicated in Alzheimer's disease.  Several strong binders to these targets are identified from the metabolite and drug datasets.  They are studied by molecular docking ~\cite{qiu_systematic_2024} and pharmocophore identification, showing promising results. Future work will aim to further validate these predictions through wet-lab experiments.

All views that are currently integrated into the MMELON architecture are ultimately derived from so-called 1D (string) or 2D (graph and image) representations~\cite{zhanggoltsgarcia2024}.   They are moderately correlated with fingerprints and contain overlapping information (Fig. \ref{fig:modelembeds}).   More expressive representations including 3D dimensional conformers~\cite{zhou2023unimol,cheng2024molecular}, and molecules in the context of binding~\cite{morrone_combining_2020,kinasedocking2024} can be readily incorporated into MMELON.  Combining insights from the molecular docking that is used to validate our approach~\cite{qiu_systematic_2024} is also a potential next step.  Besides molecule-target and property prediction, this approach can be leveraged for molecule generation and lead optimization as well.  Furthermore, while the small molecule use case is explored here, the approach itself may be readily extended to include proteins and other macromolecules.


\section{Methods}\label{methods}
\subsection{Pre-training data}\label{methods_pretrain_data}
The objective of the pre-training dataset is to develop a task-agnostic representation through self-supervision. Choosing a dataset with the right diversity and relevance to downstream tasks leads to compute-optimal richer representations. We use two main sources of molecular data. (1) PubChem \cite{kim_pubchem_2022} is a public chemical database managed by the National Center for Biotechnology Information (NCBI). It aggregates data on small molecules from over 870 sources, including information on substances, bioassays, protein targets, genes, pathways, cell lines, taxonomy, and patents. (2) ZINC22~\cite{tingle_zinc-22free_2023} is an ultra-large database containing 37 billion molecules, focusing on those with fewer than 30 heavy atoms to ensure dense coverage of “drug-like” chemical space. 

Various studies~\cite{kaplan2020scaling} indicate that the required dataset size correlates with the number of parameters used, even though the scaling laws for molecular foundation models remain to be explored fully. Furthermore, training on larger datasets tends to yield diminishing returns. For example, MolFormer was trained using a combination of PubChem and ZINC databases, encompassing a total of 1.1 billion molecules. Notably, MolFormer trained on a smaller dataset (10\% ZINC + 10\% PubChem) performed nearly as well as the model trained on a dataset ten times larger across various downstream tasks~\cite{ross_large-scale_2022}. Combining only 10\% of ZINC with 10\% of PubChem did not show reduced performance compared to training solely on ZINC. 

Our pre-training dataset comprises a collection of molecules sourced from the PubChem~\cite{kim_pubchem_2022} and ZINC22~\cite{tingle_zinc-22free_2023} chemical databases. The current size of the dataset is 200 million molecules. From PubChem, a ``drug-like" subset of 80 million molecules was curated. The curation process involved extracting the largest fragment of each molecule, removing duplicates, and filtering based on the following drug-likeness criteria:

\begin{itemize}
    \item Molecular weight $\leq 600$ Da 
    \item $\leq 5$ hydrogen bond donors
    \item $\leq 10$ hydrogen bond acceptors
    \item $\leq 10$ rotatable bonds
\end{itemize}

From the ZINC22 dataset, 120 million molecules were sampled. ZINC22 is an ultra-large database containing 37 billion molecules, focusing on those with fewer than 30 heavy atoms to ensure dense coverage of chemical space containing ``drug-like" molecules. Previous studies have extensively utilized these datasets. For example, MolFormer used the complete PubChem database combined with 1 billion molecules from ZINC, while ImageMol utilized 10\% of PubChem's data.  Figures S1-S2 illustrate the distribution of molecular properties of the pre-training dataset.

\subsection{Input representation}\label{methods_input_rep}

\paragraph{Graph Representation}
The molecular graph representation encodes a molecule as an undirected graph \( G = (V, E) \), where \( V \) represents the set of atoms as nodes and \( E \) represents the set of bonds as edges between the nodes. We employ deep graph featurization from \cite{hu_strategies_2020}. Each node \( v \in V \) is associated with a feature vector \( \mathbf{h}_v \in \mathbb{R}^d \), which encodes atom-specific properties such as atomic number, chirality, degree, formal charge, and hybridization state. Similarly, each edge \( e = (v_i, v_j) \in E \), representing a bond between atoms \( v_i \) and \( v_j \), is characterized by a feature vector \( \mathbf{h}_e \), which includes bond-specific properties such as bond type, bond direction, stereochemistry, conjugation, and whether the bond is part of a ring. The node and edge features are then embedded into 512 dimensional vectors using categorical embedding techniques, with each feature mapped to an embedding space of dimension 128. Graph Laplacian is computed over the molecular graph \( G \), yielding the eigenvalues and eigenvectors that describe the graph's structure. This additional spectral information is used as positional encoding similar to \cite{kim_pure_2022}. 

\paragraph{Image Representation}
The image representation captures the 2D visual depiction of the molecular structure, providing a spatially coherent view of atomic arrangements and functional groups. Images often highlight molecular features such as symmetry, bond angles, and ring structures, which are essential for understanding molecular shape and stereochemistry. 
The input representation for the image model in our framework follows the approach outlined in ImageMol but is reproduced here for completeness. The molecular images created using RDKit~\cite{rdkit} are processed as fixed-size pixel grids, where each pixel grid encodes local chemical environment information. This view allows the model to focus on both global molecular topology and local substructure details. The images undergo several preprocessing steps to ensure consistency and optimize the learning process. During training, data augmentation techniques are applied to the molecular images, introducing random transformations such as cropping and flipping. This augmentation process increases the model's robustness by simulating variations in the data. For validation and testing, only central cropping is applied, standardizing the image size to \(224 \times 224\) pixels without altering the original molecular structure. A normalization step follows, scaling the pixel values to a standardized range for efficient learning.

\paragraph{Text Representation}
The input representation for our Text model follows the approach outlined in MolFormer~\cite{ross_large-scale_2022}, which is a molecular transformer model that processes chemical structures represented as SMILES (Simplified Molecular Input Line Entry System) strings \cite{weininger1988smiles}. For clarity, we describe the process here.  The SMILES strings are tokenized using a custom tokenizer built upon a predefined vocabulary. The tokenizer employs a regular expression to capture key molecular substructures, atoms, bonds, and other chemical tokens, such as atoms (e.g., C, O, N), bond types (e.g., =, \#, -), and structural elements like rings and branches. The regular expression used is adapted from the approach presented in Ref. \cite{schwaller2018found}.  The tokenized SMILES strings are embedded into continuous 768 dimensional vectors.

\subsection{Architectures}\label{methods_arch}
\paragraph{Graph model}
The Graph model in our framework is based on the TokenGT architecture, which treats nodes and edges in a molecular graph as independent tokens. TokenGT provides two positional encoding schemes: orthogonal random features (ORFs) and Laplacian eigenvectors, which capture graph structure. In our implementation, we use Laplacian eigenvectors to encode positional information, as they capture the structural relationships between atoms through the graph Laplacian. The remaining layers of the model use a standard transformer architecture without graph-specific modifications. By combining explicit featurization of tokens with the self-attention mechanism of the transformer, graph model captures both local molecular features and global structural relationships without having to relearn easily computable features at the token level.

\paragraph{Image model}
The Image model is based on the ImageMol architecture, which leverages a ResNet-18 \cite{he2016deep} convolutional backbone with residual blocks for hierarchical feature extraction from molecular images. For further architectural details, please refer to Ref. \cite{zeng_accurate_2022}.

\paragraph{Text model}
The Text model is built on the MolFormer architecture, which employs a standard transformer stack to embed tokenized SMILES sequence. Rotary positional embeddings are used to encode positional information in the sequence and a linear attention mechanism~\cite{su2024roformer,katharopoulos_et_al_2020} is used within its transformer architecture.  For further architectural details, we refer the reader to Ref. \cite{ross_large-scale_2022}.

\subsection{Pre-training tasks}\label{methods_graph}

The Text model is pre-trained using a masked language modeling (MLM) task similar to Molformer. During pre-training, 15\% of the tokens in the SMILES sequences are randomly masked, and the model is trained to predict the masked tokens based on their context. The pre-training objective minimizes the cross-entropy loss between the predicted tokens and the true masked tokens. 

The Image model is pre-trained using four of the five self-supervised pretext tasks specified in the ImageMol. These tasks include: (1) Multi-Granularity Clustering Classification, which assigns pseudolabels at different levels of granularity to each molecule based on its chemical structure computed using Morgan fingerprints; (2) Molecular Rationality Discrimination, where the model distinguishes between rational and irrational molecular images; (3) Jigsaw Puzzle Prediction, where the model learns spatial relationships by solving shuffled molecular images; (4) Mask-based Contrastive Learning, which ensures that masked molecular images are consistent with their unmasked counterparts in the latent space; (5) Molecular Image Reconstruction, to enforce consistency between the image and its encoding.

The Graph model is pre-trained using three tasks designed to learn meaningful representations of molecular graphs. The first task is Masked Feature Prediction, where a portion (85\%) of the token (both atoms and bonds) features is masked, and the model is trained to predict these masked features based on the surrounding graph structure. This encourages the model to capture the local atomic environment and chemical composition. The second task is corrupted edge prediction, where 15\% of randomly selected edges have the source/edge indices changed to a different value within the graph. The model is trained to predict which edges are corrupted and which are original, based on the token features and the overall graph structure. This helps the model learn the local molecular connectivity and the chemical viability of the edges. 

\paragraph{Pre-training on topological context} The third, novel pre-training task involves predicting topological invariants of the molecular graph, specifically the Betti numbers, which are derived from simplicial homology. Betti numbers are used to describe the topological properties of a graph: \(\beta_0\) represents the number of connected components, while \(\beta_1\) captures the number of independent cycles within the graph. For each node \( v_a \), the model predicts these Betti numbers for the subgraph around that node. For example, in Fig. S7 at node \( v_a \), the graph might have \( \beta_0(S_{v_a}) = 1 \) (indicating a single connected component) and \( \beta_1(S_{v_a}) = 1 \) (indicating the presence of one cycle). In contrast, at node \( v_b \), the graph might exhibit \( \beta_0(S_{v_b}) = 2 \) (two connected components) and \( \beta_1(S_{v_b}) = 0 \) (no cycles). These Betti numbers provide both local and global structural information about the molecular graph, allowing the model to capture wider topological context at atom level representations.

\subsection{Late fusion strategies}\label{methods_agg}
\paragraph{Multi-view late fusion approach}

As first discussed in Sec. \ref{pret}, we utilize a late-fusion approach to combine Graph, Image, and Text Views. It is an attention-based approach, inspired by Ref. \cite{zhu_improving_2022} where the coefficients combining the $m^\text{th}$ view, $\alpha^m$, are the soft max outputs from an operation  on embeddings $z^m$ performed on batch $\mathcal{B}$ (Eqn. \ref{alpha_eqn}),
\begin{align}
z_i^\text{mv} &= \text{MLP} \left[ \sum\limits_{m\in\mathcal{M}} \alpha^m z^m_i \right] \label{mmf_eqn} \\ 
\alpha^m = \text{softmax}(w^m) &, \;\; w^m = \frac{1}{\mathcal{B}} \sum\limits_{i\in\mathcal{B}} \mathbf{q}^\text{T} \cdot \tanh \left(\textbf{W}\frac{z_i^m}{||z_i^m||^2} + \textbf{b} \right) \;. \label{alpha_eqn}
\end{align}
Before recovering the multi-view embedding, $z^\text{mv}$, a multi-layer perceptron is applied.  

Eqns. \ref{mmf_eqn} and \ref{alpha_eqn} form the aggregator network of the multi-view model. This network is subject to a secondary pre-training.  Aggregator pre-training yields an initialization of the weights that was found to improve robustness during fine-tuning and downstream task prediction.  The aggregator is pre-trained using an embedding reconstruction task on a set of 10M molecules randomly sampled from the dataset from which we pre-train the Graph and Text models.   At the conclusion of this secondary pre-training, it is found that the Image, Graph and Text are weighted in decreasing order, with $\alpha_\text{pret}$ values of $0.6$, $0.3$, and $0.1$, respectively.  These values do not reflect singe-view embedding quality or their contributions to specific down-stream asks after subsequent fine-tuning (see Secs. \ref{baseline_results}).  Instead, the pre-training task is related to difficulty by which a specific view can be reproduced (reconstructed) from the combined embedding and is consistent with the observation that Image is less correlated to the Graph and Text views (see Sec. \ref{fig:modelembeds}C).

\paragraph{Alternative late fusion strategies}
Before choosing our multi-view model architecture, we  explored several other late fusion strategies that combine the representations from the Graph, Text, and Image models. These strategies differ in how the inputs are prepared for the gating network. The gating network computes weights for each view, and the final output is a weighted aggregation of the view-specific representations.

1. Projected Gating: The representations from each model are projected into a lower-dimensional space. Let \( d_i \) denote the dimensionality of the output from modality \( i \), where \( i \in \{1, 2, 3\} \) corresponds to the graph, text, and image models. The total input dimension to the gating network is:
   \[
   D_{\text{projected}} = \min(d_1, d_2, d_3) \times 3
   \]
The concatenated projected embeddings are passed into the gating network, which computes the weights \( w_i \) for each modality.

2. Unprojected Gating: In this scheme, the original (unprojected) representations from the models are concatenated and used as input to the gating network. The input dimension in this case is the sum of the dimensions of the individual outputs:
   \[
   D_{\text{unprojected}} = d_1 + d_2 + d_3
   \]
The gating network then computes the weights based on this concatenated unprojected representation. In both the schemes above, we compute weighted concatenation of the model output. 
\[
z_{\text{final}} =  w_1 z_1 \oplus w_2 z_2 \oplus w_3 z_3 
\]
3. Projected Gating With Feature Addition: This scheme the model outputs are first projected and concatenated, and the gating network computes the weights from this combined representation. The same aggregation process follows, where weights \( w_i \) are applied to the corresponding expert outputs.

In this scheme, the final representation \( z_{\text{final}} \) is computed as a weighted sum:
\[
z_{\text{final}} = \sum_{i=1}^{3} w_i z_i
\]
where \( w_i \) are the modality-specific weights and \( z_i \) are the corresponding representations. Additionally, for interpretability, the average weights across the batch can be computed:
\[
\bar{w}_i = \frac{1}{N} \sum_{n=1}^{N} w_i^{(n)}
\]
where \( N \) is the batch size.

The performance on a subset of MoleculeNet tasks for alternate schemes 1, 2, and 3 is compared against that of the multi-view model in Tables S7-S8.  Overall, the chosen multi-view architecture is the top performer.

\subsection{Drug-target interaction model} \label{methods_dti}
Ref. \cite{gorantla_proteins_2024} proposed a sequence-based framework to investigate the binding affinities of proteins and ligands whilst experimenting with various embedding types for this purpose. From this study, we adapted their model for representing the protein structure that uses convolutional neural network-based encodings obtained from ESM-1b~\cite{Rao2020.12.15.422761}. We combined these CNN-based protein embeddings with ligand embeddings obtained from our models. The combined embeddings were then passed through a prediction head, which was based on the prediction head of Ref.  \cite{gorantla_proteins_2024}.  We used 64-dimensional protein and ligand encodings which are then concatenated to form a 128-dimensional embedding.  The authors of Ref. \cite{gorantla_proteins_2024} used double this size.
We experimented with combining the protein embeddings with ligand embeddings obtained from various modalities of representation developed here: Image, Text, Graph, and multi-view. 

During training, the protein, single- and multi- view model weights were unfrozen gradually using the Xavier method. We used a batch size of 64 and a learning rate of 0.0005 for all models except for the Text model, for which we used a batch size of 128 and a learning rate of 0.00001. We trained for 2000 epochs, as suggested in Ref  \cite{gorantla_proteins_2024}, and used the best performing model on the validation set to predict the test set. We used six-fold cross-validation using one-sixth of the data for validation, one-sixth for testing, and the rest for training. The results were then averaged across the six cross-validation folds.

\subsection{Statistical analysis of select classification models} \label{methods_statistics}
We consider the fine-tuned output of the three single-view and multi-view models in single-task classification and regression tasks and aim to perform comparative analyses to uncover distinct characterizations of the models' predictions. For classification tasks, binarization was performed on outputs by applying a threshold of 0.5 to the sigmoid mapped scores.  $F1$ scores are employed to evaluate the agreement between models.  $F1$ scores compare binarized predictions typical of classifications, measuring how well the second model's predicted activities predict the first model's predicted activities.  Note that this measure is not symmetric.  For regression tasks, the model outputs for continuous prediction results are compared using the Pearson correlation coefficient, with a $\chi^2$ p-value estimator.

\subsection{Fine-tuning protocols}\label{methods_ft}

The choice of data splitting into training, validation and test sets greatly impacts model performance.  For example, The size-ordered scaffold split clusters molecules according to their Murko scaffold~\cite{bemis_properties_1996} and assigns the clusters to splits in size-order. This is to be compared with other variations of scaffold split, which  cluster the dataset in the same manner but assign clusters to train, validation, and test partitions randomly or `balanced' (see, e.g. Ref. ~\cite{zeng_accurate_2022}). The balanced scaffold split assigns clusters large relative to the test set size to the training and other clusters randomly. Scaffold splits are generally understood to be more appropriate and harder than random splits, though the latter are still often utilized in benchmarks~\cite{MoleculeNet}.  Following Ref. ~\cite{zeng_accurate_2022} size-ordered scaffold and were used for MoleculeNet and  balanced scaffold splits for CYP and GPCR tasks with an 0.8/0.1/0.1 train/validation/test splitting.  For the ComputationalADME datasets it is noted that scaffold clustering yields a large number of singletons~\cite{computationalADME2023} so a random split (0.7/0.1/0.2 train/validation/test) was employed.

Hyper-parameter optimization of the fine-tuning process is important to ensure quality outputs. Optimization was first explored using Ray-Tune~\cite{liaw2018tune}.   Based on an exploration of our multi-view model across a subset of MoleculeNet, a set of parameters for AdamW and regularization (e.g. weight-decay) were chosen.  Linear and MLP prediction heads were also explored.  The learning rate was then varied for individual MoleculeNet tasks for Image, Graph, Text in the range of 1e-5 to 1e-3).  In the case of the multi-view model, the presence of multiple sub-networks may call for a multiple optimizer fine-tuning strategy that employs up to four learning rates in the aforementioned range.  For CYP, ComputationalADME and GPCR models, a learning rate was chosen across each set of tasks. The highest average validation performer across 5 seeds was chosen as the model instance to report the held-out (test) set results.

In order to facilitate the plot in  Fig. \ref{fig:molnet_scaffold} we scale the regression metrics according to the following formula $m^\prime = 1- (m-m_0)/s$ where $m$ is the unscaled metric and $m_0$ and $s$ are an offset and scaling factor, respectively.  For ESOL, FreeSolv, Lipophilicity and QM7 the parameters $(m_0,s)$ are $(0.7,1.0)$, $(1.5,5.0)$, $(0.50,1.0)$, and $(55.0,100.0)$, respectively.  For the six ComputationalADME datasets, parameters $(0.30,1.0)$ are used.

\subsection{Genetics and multi-omics analysis}\label{methods_evidence}
The genetics-informed GPCRs prioritized by Mendelian randomization (MR) were retrieved from our previous study \cite{qiu_systematic_2024}. Briefly, 408 GPCRs were tested and publicly available 4 cis-eQTL (Expression quantitative trait locus) datasets of human cortex region and 3 publicly available AD GWAS summary statistic datasets were used to test the causal effects of GPCRs to AD.
Multi-omics data of 88 bulk and single cell RNA-seq transcriptome or proteome datasets were used we previously compiled (available at AlzGPS: https://alzgps.lerner.ccf.org/) \cite{Zhou2021AlzGPS}. Briefly, differential expression analysis was performed on microarray, bulk RNA-seq, and single-cell/nucleus (sc/sn) RNA-seq datasets. The threshold of differentially expressed genes (DEGs) were defined with adjusted p-value (q) \textless 0.05 and $|log_2(\mathrm{FC})| \geq 0.25$.

\subsection{Molecular docking simulations}
\label{methods_docking}
3D structure of CYP2C9 complexed with a drug Lorsatan was retrieved from Protein Data Bank (PDB ID: 8VX0)\cite{Parikh2024Structural}. The top bioactive molecules (CHEBL1466628, CHEMBL1353420, and CHEMBL1742011) were prioritized from our classification model.  Top 10 drugs or metabolites of AD genetics-informed FPR1 and AD multi-omics-informed ADA2A were inspected. 3D structures of FPR1 or ADA2A were retrieved from AlphaFold2 database\cite{Jumper2021Highly}. The preparation of 2D structures of small molecules and 3D structures of proteins refer to our previous study\cite{qiu_systematic_2024}. Molecular docking was processed by AutoDock Vina (version 1.1.2)\cite{Eberhardt2021AutoDock}. Heme was kept for CYP2C9 molecular docking. The top 10 binding modes were investigated, and the best binding poses were selected for further analysis. Detailed binding analysis were conducted by using PyMOL (version 3.0.3).

\backmatter

\bmhead{Code Availability}
Our Multi-View model is available on GitHub (\url{https://github.com/BiomedSciAI/biomed-multi-view}) and Hugging Face (\url{https://Huggingface.co/ibm-research/biomed.sm.mv-te-84m}). 

\bmhead{Acknowledgments}
We acknowledge Viatcheslav Gurev, William Ogallo, Sekou Remy, Mohamed Ghalwash, Leili Zhang, Wendy Cornell, James Kozloski, Michal Rosen-Zvi and Ajay Royyuru for insightful discussions.


\bibliography{article}

\begin{thebibliography}{10}
\expandafter\ifx\csname url\endcsname\relax
  \def\url#1{\burl{#1}}\fi
\expandafter\ifx\csname urlprefix\endcsname\relax\def\urlprefix{URL }\fi
\providecommand{\bibinfo}[2]{#2}
\providecommand{\eprint}[2][]{\url{#2}}
\providecommand{\doi}[1]{\url{https://doi.org/#1}}
\bibcommenthead

\bibitem{hughes_principles_2011}
\bibinfo{author}{Hughes, J.}, \bibinfo{author}{Rees, S.}, \bibinfo{author}{Kalindjian, S.} \& \bibinfo{author}{Philpott, K.}
\newblock \bibinfo{title}{Principles of early drug discovery}.
\newblock \emph{\bibinfo{journal}{British Journal of Pharmacology}} \textbf{\bibinfo{volume}{162}}, \bibinfo{pages}{1239--1249} (\bibinfo{year}{2011}).
\newblock \urlprefix\url{https://doi.org/10.1111/j.1476-5381.2010.01127.x}.
\newblock \bibinfo{note}{Publisher: John Wiley \& Sons, Ltd}.

\bibitem{zeng_deep_2022}
\bibinfo{author}{Zeng, X.} \emph{et~al.}
\newblock \bibinfo{title}{Deep generative molecular design reshapes drug discovery}.
\newblock \emph{\bibinfo{journal}{Cell Reports Medicine}} \textbf{\bibinfo{volume}{3}} (\bibinfo{year}{2022}).
\newblock \urlprefix\url{https://doi.org/10.1016/j.xcrm.2022.100794}.
\newblock \bibinfo{note}{Publisher: Elsevier}.

\bibitem{cheng_artificial_2024}
\bibinfo{author}{Cheng, F.} \emph{et~al.}
\newblock \bibinfo{title}{Artificial intelligence and open science in discovery of disease-modifying medicines for {Alzheimer}’s disease}.
\newblock \emph{\bibinfo{journal}{Cell Reports Medicine}} \textbf{\bibinfo{volume}{5}} (\bibinfo{year}{2024}).
\newblock \urlprefix\url{https://doi.org/10.1016/j.xcrm.2023.101379}.
\newblock \bibinfo{note}{Publisher: Elsevier}.

\bibitem{butkiewicz_benchmarking_2013}
\bibinfo{author}{Butkiewicz, M.} \emph{et~al.}
\newblock \bibinfo{title}{Benchmarking {Ligand}-{Based} {Virtual} {High}-{Throughput} {Screening} with the {PubChem} {Database}}.
\newblock \emph{\bibinfo{journal}{Molecules}} \textbf{\bibinfo{volume}{18}}, \bibinfo{pages}{735--756} (\bibinfo{year}{2013}).

\bibitem{vamathevan_applications_2019}
\bibinfo{author}{Vamathevan, J.} \emph{et~al.}
\newblock \bibinfo{title}{Applications of machine learning in drug discovery and development}.
\newblock \emph{\bibinfo{journal}{Nature Reviews Drug Discovery}} \textbf{\bibinfo{volume}{18}}, \bibinfo{pages}{463--477} (\bibinfo{year}{2019}).
\newblock \urlprefix\url{https://doi.org/10.1038/s41573-019-0024-5}.

\bibitem{muratov_qsar_2020}
\bibinfo{author}{Muratov, E.~N.} \emph{et~al.}
\newblock \bibinfo{title}{{QSAR} without borders}.
\newblock \emph{\bibinfo{journal}{Chemical Society Reviews}} \textbf{\bibinfo{volume}{49}}, \bibinfo{pages}{3525--3564} (\bibinfo{year}{2020}).
\newblock \urlprefix\url{http://dx.doi.org/10.1039/D0CS00098A}.
\newblock \bibinfo{note}{Publisher: The Royal Society of Chemistry}.

\bibitem{wigh_review_2022}
\bibinfo{author}{Wigh, D.~S.}, \bibinfo{author}{Goodman, J.~M.} \& \bibinfo{author}{Lapkin, A.~A.}
\newblock \bibinfo{title}{A review of molecular representation in the age of machine learning}.
\newblock \emph{\bibinfo{journal}{WIREs Computational Molecular Science}} \textbf{\bibinfo{volume}{12}}, \bibinfo{pages}{e1603} (\bibinfo{year}{2022}).
\newblock \urlprefix\url{https://doi.org/10.1002/wcms.1603}.
\newblock \bibinfo{note}{Publisher: John Wiley \& Sons, Ltd}.

\bibitem{zhao2024surveylargelanguagemodels}
\bibinfo{author}{Zhao, W.~X.} \emph{et~al.}
\newblock \bibinfo{title}{A survey of large language models} (\bibinfo{year}{2024}).
\newblock \urlprefix\url{https://arxiv.org/abs/2303.18223}.
\newblock \eprint{2303.18223}.

\bibitem{chithrananda_chemberta_2020}
\bibinfo{author}{Chithrananda, S.}, \bibinfo{author}{Grand, G.} \& \bibinfo{author}{Ramsundar, B.}
\newblock \bibinfo{title}{{ChemBERTa}: {Large}-{Scale} {Self}-{Supervised} {Pretraining} for {Molecular} {Property} {Prediction}} (\bibinfo{year}{2020}).
\newblock \urlprefix\url{http://arxiv.org/abs/2010.09885}.
\newblock \bibinfo{note}{ArXiv:2010.09885 [physics, q-bio]}.

\bibitem{chilingaryan_bartsmiles_2022}
\bibinfo{author}{Chilingaryan, G.} \emph{et~al.}
\newblock \bibinfo{title}{{BARTSmiles}: {Generative} {Masked} {Language} {Models} for {Molecular} {Representations}} (\bibinfo{year}{2022}).
\newblock \urlprefix\url{http://arxiv.org/abs/2211.16349}.
\newblock \bibinfo{note}{ArXiv:2211.16349 [cs, q-bio]}.

\bibitem{ross_large-scale_2022}
\bibinfo{author}{Ross, J.} \emph{et~al.}
\newblock \bibinfo{title}{Large-scale chemical language representations capture molecular structure and properties}.
\newblock \emph{\bibinfo{journal}{Nature Machine Intelligence}} \textbf{\bibinfo{volume}{4}}, \bibinfo{pages}{1256--1264} (\bibinfo{year}{2022}).
\newblock \urlprefix\url{https://doi.org/10.1038/s42256-022-00580-7}.

\bibitem{atz_geometric_2021}
\bibinfo{author}{Atz, K.}, \bibinfo{author}{Grisoni, F.} \& \bibinfo{author}{Schneider, G.}
\newblock \bibinfo{title}{Geometric deep learning on molecular representations}.
\newblock \emph{\bibinfo{journal}{Nature Machine Intelligence}} \textbf{\bibinfo{volume}{3}}, \bibinfo{pages}{1023--1032} (\bibinfo{year}{2021}).
\newblock \urlprefix\url{https://www.nature.com/articles/s42256-021-00418-8}.

\bibitem{carhart_atom_1985}
\bibinfo{author}{Carhart, R.~E.}, \bibinfo{author}{Smith, D.~H.} \& \bibinfo{author}{Venkataraghavan, R.}
\newblock \bibinfo{title}{Atom pairs as molecular features in structure-activity studies: definition and applications}.
\newblock \emph{\bibinfo{journal}{Journal of Chemical Information and Computer Sciences}} \textbf{\bibinfo{volume}{25}}, \bibinfo{pages}{64--73} (\bibinfo{year}{1985}).
\newblock \urlprefix\url{https://doi.org/10.1021/ci00046a002}.
\newblock \bibinfo{note}{Publisher: American Chemical Society}.

\bibitem{durant_reoptimization_2002}
\bibinfo{author}{Durant, J.~L.}, \bibinfo{author}{Leland, B.~A.}, \bibinfo{author}{Henry, D.~R.} \& \bibinfo{author}{Nourse, J.~G.}
\newblock \bibinfo{title}{Reoptimization of {MDL} {Keys} for {Use} in {Drug} {Discovery}}.
\newblock \emph{\bibinfo{journal}{Journal of Chemical Information and Computer Sciences}} \textbf{\bibinfo{volume}{42}}, \bibinfo{pages}{1273--1280} (\bibinfo{year}{2002}).
\newblock \urlprefix\url{https://doi.org/10.1021/ci010132r}.
\newblock \bibinfo{note}{Publisher: American Chemical Society}.

\bibitem{rogers_extended-connectivity_2010}
\bibinfo{author}{Rogers, D.} \& \bibinfo{author}{Hahn, M.}
\newblock \bibinfo{title}{Extended-{Connectivity} {Fingerprints}}.
\newblock \emph{\bibinfo{journal}{Journal of Chemical Information and Modeling}} \textbf{\bibinfo{volume}{50}}, \bibinfo{pages}{742--754} (\bibinfo{year}{2010}).
\newblock \urlprefix\url{https://doi.org/10.1021/ci100050t}.
\newblock \bibinfo{note}{Publisher: American Chemical Society}.

\bibitem{oboyle_comparing_2016}
\bibinfo{author}{O’Boyle, N.~M.} \& \bibinfo{author}{Sayle, R.~A.}
\newblock \bibinfo{title}{Comparing structural fingerprints using a literature-based similarity benchmark}.
\newblock \emph{\bibinfo{journal}{Journal of Cheminformatics}} \textbf{\bibinfo{volume}{8}}, \bibinfo{pages}{36} (\bibinfo{year}{2016}).
\newblock \urlprefix\url{https://doi.org/10.1186/s13321-016-0148-0}.

\bibitem{rdkit}
\bibinfo{author}{Landrum, G.}
\newblock \bibinfo{title}{Rdkit: Open-source cheminformatics}.
\newblock \urlprefix\url{http://www.rdkit.org}.

\bibitem{NIPS2015_f9be311e}
\bibinfo{author}{Duvenaud, D.~K.} \emph{et~al.}
\newblock \bibinfo{editor}{Cortes, C.}, \bibinfo{editor}{Lawrence, N.}, \bibinfo{editor}{Lee, D.}, \bibinfo{editor}{Sugiyama, M.} \& \bibinfo{editor}{Garnett, R.} (eds) \emph{\bibinfo{title}{Convolutional networks on graphs for learning molecular fingerprints}}.
\newblock (eds \bibinfo{editor}{Cortes, C.}, \bibinfo{editor}{Lawrence, N.}, \bibinfo{editor}{Lee, D.}, \bibinfo{editor}{Sugiyama, M.} \& \bibinfo{editor}{Garnett, R.}) \emph{\bibinfo{booktitle}{Advances in Neural Information Processing Systems}}, Vol.~\bibinfo{volume}{28} (\bibinfo{publisher}{Curran Associates, Inc.}, \bibinfo{year}{2015}).
\newblock \urlprefix\url{https://proceedings.neurips.cc/paper_files/paper/2015/file/f9be311e65d81a9ad8150a60844bb94c-Paper.pdf}.

\bibitem{altae-tran_low_2017}
\bibinfo{author}{Altae-Tran, H.}, \bibinfo{author}{Ramsundar, B.}, \bibinfo{author}{Pappu, A.~S.} \& \bibinfo{author}{Pande, V.}
\newblock \bibinfo{title}{Low {Data} {Drug} {Discovery} with {One}-{Shot} {Learning}}.
\newblock \emph{\bibinfo{journal}{ACS Central Science}} \textbf{\bibinfo{volume}{3}}, \bibinfo{pages}{283--293} (\bibinfo{year}{2017}).
\newblock \urlprefix\url{https://doi.org/10.1021/acscentsci.6b00367}.
\newblock \bibinfo{note}{Publisher: American Chemical Society}.

\bibitem{morrone_combining_2020}
\bibinfo{author}{Morrone, J.~A.}, \bibinfo{author}{Weber, J.~K.}, \bibinfo{author}{Huynh, T.}, \bibinfo{author}{Luo, H.} \& \bibinfo{author}{Cornell, W.~D.}
\newblock \bibinfo{title}{Combining docking pose rank and structure with deep learning improves protein–ligand binding mode prediction over a baseline docking approach}.
\newblock \emph{\bibinfo{journal}{Journal of chemical information and modeling}} \textbf{\bibinfo{volume}{60}}, \bibinfo{pages}{4170--4179} (\bibinfo{year}{2020}).
\newblock \bibinfo{note}{Publisher: American Chemical Society}.

\bibitem{wieder_compact_2020}
\bibinfo{author}{Wieder, O.} \emph{et~al.}
\newblock \bibinfo{title}{A compact review of molecular property prediction with graph neural networks}.
\newblock \emph{\bibinfo{journal}{Drug Discovery Today: Technologies}} \textbf{\bibinfo{volume}{37}}, \bibinfo{pages}{1--12} (\bibinfo{year}{2020}).
\newblock \urlprefix\url{https://www.sciencedirect.com/science/article/pii/S1740674920300305}.

\bibitem{kim_pure_2022}
\bibinfo{author}{Kim, J.} \emph{et~al.}
\newblock \bibinfo{title}{Pure {Transformers} are {Powerful} {Graph} {Learners}} (\bibinfo{year}{2022}).
\newblock \urlprefix\url{http://arxiv.org/abs/2207.02505}.
\newblock \bibinfo{note}{ArXiv:2207.02505 [cs]}.

\bibitem{muller_attending_2023}
\bibinfo{author}{Müller, L.}, \bibinfo{author}{Galkin, M.}, \bibinfo{author}{Morris, C.} \& \bibinfo{author}{Rampášek, L.}
\newblock \bibinfo{title}{Attending to {Graph} {Transformers}} (\bibinfo{year}{2023}).
\newblock \urlprefix\url{http://arxiv.org/abs/2302.04181}.
\newblock \bibinfo{note}{ArXiv:2302.04181 [cs]}.

\bibitem{goh2017chemceptiondeepneuralnetwork}
\bibinfo{author}{Goh, G.~B.}, \bibinfo{author}{Siegel, C.}, \bibinfo{author}{Vishnu, A.}, \bibinfo{author}{Hodas, N.~O.} \& \bibinfo{author}{Baker, N.}
\newblock \bibinfo{title}{Chemception: A deep neural network with minimal chemistry knowledge matches the performance of expert-developed qsar/qspr models} (\bibinfo{year}{2017}).
\newblock \urlprefix\url{https://arxiv.org/abs/1706.06689}.
\newblock \eprint{1706.06689}.

\bibitem{zeng_accurate_2022}
\bibinfo{author}{Zeng, X.} \emph{et~al.}
\newblock \bibinfo{title}{Accurate prediction of molecular properties and drug targets using a self-supervised image representation learning framework}.
\newblock \emph{\bibinfo{journal}{Nature Machine Intelligence}} \textbf{\bibinfo{volume}{4}}, \bibinfo{pages}{1004--1016} (\bibinfo{year}{2022}).
\newblock \urlprefix\url{https://www.nature.com/articles/s42256-022-00557-6}.

\bibitem{liu_pre-training_2022}
\bibinfo{author}{Liu, S.} \emph{et~al.}
\newblock \bibinfo{title}{Pre-training {Molecular} {Graph} {Representation} with {3D} {Geometry}} (\bibinfo{year}{2022}).
\newblock \urlprefix\url{http://arxiv.org/abs/2110.07728}.
\newblock \bibinfo{note}{ArXiv:2110.07728 [cs, eess, q-bio]}.

\bibitem{zhu_unified_2022}
\bibinfo{author}{Zhu, J.} \emph{et~al.}
\newblock \bibinfo{title}{Unified {2D} and {3D} {Pre}-{Training} of {Molecular} {Representations}} \bibinfo{pages}{2626--2636} (\bibinfo{year}{2022}).
\newblock \urlprefix\url{https://dl.acm.org/doi/10.1145/3534678.3539368}.

\bibitem{zhu_improving_2022}
\bibinfo{author}{Zhu, Y.} \emph{et~al.}
\newblock \bibinfo{title}{Improving {Molecular} {Pretraining} with {Complementary} {Featurizations}} (\bibinfo{year}{2022}).
\newblock \urlprefix\url{http://arxiv.org/abs/2209.15101}.
\newblock \bibinfo{note}{ArXiv:2209.15101 [physics, q-bio]}.

\bibitem{du_fusing_2023}
\bibinfo{author}{Du, W.} \emph{et~al.}
\newblock \bibinfo{title}{Fusing {2D} and {3D} molecular graphs as unambiguous molecular descriptors for conformational and chiral stereoisomers}.
\newblock \emph{\bibinfo{journal}{Briefings in Bioinformatics}} \textbf{\bibinfo{volume}{24}}, \bibinfo{pages}{bbac560} (\bibinfo{year}{2023}).
\newblock \urlprefix\url{https://doi.org/10.1093/bib/bbac560}.

\bibitem{bmfm_blog}
\bibinfo{author}{Hu, J.} \emph{et~al.}
\newblock \bibinfo{title}{Biomedical foundation models} (\bibinfo{year}{2023}).
\newblock \urlprefix\url{https://research.ibm.com/projects/biomedical-foundation-models}.

\bibitem{kim_pubchem_2022}
\bibinfo{author}{Kim, S.} \emph{et~al.}
\newblock \bibinfo{title}{{PubChem} 2023 update}.
\newblock \emph{\bibinfo{journal}{Nucleic Acids Research}} \textbf{\bibinfo{volume}{51}}, \bibinfo{pages}{D1373--D1380} (\bibinfo{year}{2022}).
\newblock \urlprefix\url{https://doi.org/10.1093/nar/gkac956}.
\newblock \bibinfo{note}{\_eprint: https://academic.oup.com/nar/article-pdf/51/D1/D1373/48441598/gkac956.pdf}.

\bibitem{tingle_zinc-22free_2023}
\bibinfo{author}{Tingle, B.~I.} \emph{et~al.}
\newblock \bibinfo{title}{Zinc-22-a free multi-billion-scale database of tangible compounds for ligand discoery}.
\newblock \emph{\bibinfo{journal}{Journal of Chemical Information and Modeling}} \textbf{\bibinfo{volume}{63}}, \bibinfo{pages}{1166--1176} (\bibinfo{year}{2023}).
\newblock \urlprefix\url{https://doi.org/10.1021/acs.jcim.2c01253}.
\newblock \bibinfo{note}{Publisher: American Chemical Society}.

\bibitem{cheng_classification_2011}
\bibinfo{author}{Cheng, F.} \emph{et~al.}
\newblock \bibinfo{title}{Classification of {Cytochrome} {P450} {Inhibitors} and {Noninhibitors} {Using} {Combined} {Classifiers}}.
\newblock \emph{\bibinfo{journal}{Journal of Chemical Information and Modeling}} \textbf{\bibinfo{volume}{51}}, \bibinfo{pages}{996--1011} (\bibinfo{year}{2011}).
\newblock \urlprefix\url{https://doi.org/10.1021/ci200028n}.
\newblock \bibinfo{note}{Publisher: American Chemical Society}.

\bibitem{MoleculeNet}
\bibinfo{author}{Wu, Z.} \emph{et~al.}
\newblock \bibinfo{title}{Moleculenet: a benchmark for molecular machine learning}.
\newblock \emph{\bibinfo{journal}{Chem. Sci.}} \textbf{\bibinfo{volume}{9}}, \bibinfo{pages}{513--530} (\bibinfo{year}{2018}).
\newblock \urlprefix\url{http://dx.doi.org/10.1039/C7SC02664A}.

\bibitem{davis_comprehensive_2011}
\bibinfo{author}{Davis, M.~I.} \emph{et~al.}
\newblock \bibinfo{title}{Comprehensive analysis of kinase inhibitor selectivity}.
\newblock \emph{\bibinfo{journal}{Nature Biotechnology}} \textbf{\bibinfo{volume}{29}}, \bibinfo{pages}{1046--1051} (\bibinfo{year}{2011}).
\newblock \urlprefix\url{https://doi.org/10.1038/nbt.1990}.

\bibitem{2024AD}
\bibinfo{title}{2024 alzheimer's disease facts and figures}.
\newblock \emph{\bibinfo{journal}{Alzheimer's \& dementia : the journal of the Alzheimer's Association}} \textbf{\bibinfo{volume}{20}}, \bibinfo{pages}{3708--3821} (\bibinfo{year}{2024}).
\newblock \bibinfo{note}{[Online; accessed 2024-12-02]}.

\bibitem{Cummings2024Alzheimer}
\bibinfo{author}{Cummings, J.} \emph{et~al.}
\newblock \bibinfo{title}{Alzheimer's disease drug development pipeline: 2024}.
\newblock \emph{\bibinfo{journal}{Alzheimer's \& dementia (New York, N. Y.)}} \textbf{\bibinfo{volume}{10}}, \bibinfo{pages}{e12465} (\bibinfo{year}{2024}).
\newblock \bibinfo{note}{[Online; accessed 2024-12-02]}.

\bibitem{Qiu2024Artificial}
\bibinfo{author}{Qiu, Y.} \& \bibinfo{author}{Cheng, F.}
\newblock \bibinfo{title}{Artificial intelligence for drug discovery and development in alzheimer's disease}.
\newblock \emph{\bibinfo{journal}{Current opinion in structural biology}} \textbf{\bibinfo{volume}{85}}, \bibinfo{pages}{102776} (\bibinfo{year}{2024}).
\newblock \bibinfo{note}{[Online; accessed 2024-12-02]}.

\bibitem{Xu2022Interpretable}
\bibinfo{author}{Xu, J.} \emph{et~al.}
\newblock \bibinfo{title}{Interpretable deep learning translation of gwas and multi-omics findings to identify pathobiology and drug repurposing in alzheimer's disease}.
\newblock \emph{\bibinfo{journal}{Cell reports}} \textbf{\bibinfo{volume}{41}}, \bibinfo{pages}{111717} (\bibinfo{year}{2022}).
\newblock \bibinfo{note}{[Online; accessed 2024-12-02]}.

\bibitem{Thathiah2011role}
\bibinfo{author}{Thathiah, A.} \& \bibinfo{author}{De~Strooper, B.}
\newblock \bibinfo{title}{The role of g protein-coupled receptors in the pathology of alzheimer's disease}.
\newblock \emph{\bibinfo{journal}{Nature reviews. Neuroscience}} \textbf{\bibinfo{volume}{12}}, \bibinfo{pages}{73--87} (\bibinfo{year}{2011}).
\newblock \bibinfo{note}{[Online; accessed 2024-12-02]}.

\bibitem{yang2024deep}
\bibinfo{author}{Yang, Y.} \emph{et~al.}
\newblock \bibinfo{title}{A deep learning framework combining molecular image and protein structural representations identifies candidate drugs for pain}.
\newblock \emph{\bibinfo{journal}{Cell Reports Methods}} \textbf{\bibinfo{volume}{4}} (\bibinfo{year}{2024}).

\bibitem{yang2024deepm}
\bibinfo{author}{Yang, Y.} \emph{et~al.}
\newblock \bibinfo{editor}{Partin-Vaisband, I.}, \bibinfo{editor}{Katkoori, S.}, \bibinfo{editor}{Peng, L.}, \bibinfo{editor}{Vaisband, B.} \& \bibinfo{editor}{Nikoubin, T.} (eds) \emph{\bibinfo{title}{A deep multimodal representation learning framework for accurate molecular properties prediction}}.
\newblock (eds \bibinfo{editor}{Partin-Vaisband, I.}, \bibinfo{editor}{Katkoori, S.}, \bibinfo{editor}{Peng, L.}, \bibinfo{editor}{Vaisband, B.} \& \bibinfo{editor}{Nikoubin, T.}) \emph{\bibinfo{booktitle}{Proceedings of the Great Lakes Symposium on VLSI 2024}}, GLSVLSI '24, \bibinfo{pages}{760--765} (\bibinfo{publisher}{Association for Computing Machinery}, \bibinfo{address}{New York, NY, USA}, \bibinfo{year}{2024}).
\newblock \urlprefix\url{https://doi.org/10.1145/3649476.3660377}.

\bibitem{chembl2011}
\bibinfo{author}{Gaulton, A.} \emph{et~al.}
\newblock \bibinfo{title}{{ChEMBL: a large-scale bioactivity database for drug discovery}}.
\newblock \emph{\bibinfo{journal}{Nucleic Acids Research}} \textbf{\bibinfo{volume}{40}}, \bibinfo{pages}{D1100--D1107} (\bibinfo{year}{2011}).
\newblock \urlprefix\url{https://doi.org/10.1093/nar/gkr777}.

\bibitem{glass2015}
\bibinfo{author}{Chan, W. K.~B.} \emph{et~al.}
\newblock \bibinfo{title}{{GLASS: a comprehensive database for experimentally validated GPCR-ligand associations}}.
\newblock \emph{\bibinfo{journal}{Bioinformatics}} \textbf{\bibinfo{volume}{31}}, \bibinfo{pages}{3035--3042} (\bibinfo{year}{2015}).
\newblock \urlprefix\url{https://doi.org/10.1093/bioinformatics/btv302}.

\bibitem{drugbank2018}
\bibinfo{author}{Wishart, D.~S.} \emph{et~al.}
\newblock \bibinfo{title}{{DrugBank 5.0: a major update to the DrugBank database for 2018}}.
\newblock \emph{\bibinfo{journal}{Nucleic Acids Research}} \textbf{\bibinfo{volume}{46}}, \bibinfo{pages}{D1074--D1082} (\bibinfo{year}{2017}).
\newblock \urlprefix\url{https://doi.org/10.1093/nar/gkx1037}.

\bibitem{LIPINSKI20013}
\bibinfo{author}{Lipinski, C.~A.}, \bibinfo{author}{Lombardo, F.}, \bibinfo{author}{Dominy, B.~W.} \& \bibinfo{author}{Feeney, P.~J.}
\newblock \bibinfo{title}{Experimental and computational approaches to estimate solubility and permeability in drug discovery and development settings}.
\newblock \emph{\bibinfo{journal}{Advanced Drug Delivery Reviews}} \textbf{\bibinfo{volume}{46}}, \bibinfo{pages}{3--26} (\bibinfo{year}{2001}).
\newblock \urlprefix\url{https://www.sciencedirect.com/science/article/pii/S0169409X00001290}.

\bibitem{liang_foundations_2023}
\bibinfo{author}{Liang, P.~P.}, \bibinfo{author}{Zadeh, A.} \& \bibinfo{author}{Morency, L.-P.}
\newblock \bibinfo{title}{Foundations and {Trends} in {Multimodal} {Machine} {Learning}: {Principles}, {Challenges}, and {Open} {Questions}} (\bibinfo{year}{2023}).
\newblock \urlprefix\url{http://arxiv.org/abs/2209.03430}.
\newblock \bibinfo{note}{ArXiv:2209.03430 [cs]}.

\bibitem{nilakantan_topological_1987}
\bibinfo{author}{Nilakantan, R.}, \bibinfo{author}{Bauman, N.}, \bibinfo{author}{Dixon, J.~S.} \& \bibinfo{author}{Venkataraghavan, R.}
\newblock \bibinfo{title}{Topological torsion: a new molecular descriptor for {SAR} applications. {Comparison} with other descriptors}.
\newblock \emph{\bibinfo{journal}{Journal of Chemical Information and Computer Sciences}} \textbf{\bibinfo{volume}{27}}, \bibinfo{pages}{82--85} (\bibinfo{year}{1987}).
\newblock \urlprefix\url{https://doi.org/10.1021/ci00054a008}.
\newblock \bibinfo{note}{Publisher: American Chemical Society}.

\bibitem{computationalADME2023}
\bibinfo{author}{Fang, C.} \emph{et~al.}
\newblock \bibinfo{title}{Prospective validation of machine learning algorithms for absorption, distribution, metabolism, and excretion prediction: An industrial perspective}.
\newblock \emph{\bibinfo{journal}{Journal of Chemical Information and Modeling}} \textbf{\bibinfo{volume}{63}}, \bibinfo{pages}{3263--3274} (\bibinfo{year}{2023}).
\newblock \urlprefix\url{https://doi.org/10.1021/acs.jcim.3c00160}.
\newblock \bibinfo{note}{PMID: 37216672}.

\bibitem{fey2019fastgraphrepresentationlearning}
\bibinfo{author}{Fey, M.} \& \bibinfo{author}{Lenssen, J.~E.}
\newblock \bibinfo{title}{Fast graph representation learning with pytorch geometric} (\bibinfo{year}{2019}).
\newblock \urlprefix\url{https://arxiv.org/abs/1903.02428}.
\newblock \eprint{1903.02428}.

\bibitem{Zhao2021Cytochrome}
\bibinfo{author}{Zhao, M.} \emph{et~al.}
\newblock \bibinfo{title}{Cytochrome p450 enzymes and drug metabolism in humans}.
\newblock \emph{\bibinfo{journal}{International Journal of Molecular Sciences}} \textbf{\bibinfo{volume}{22}}, \bibinfo{pages}{12808} (\bibinfo{year}{2021}).
\newblock \urlprefix\url{http://dx.doi.org/10.3390/ijms222312808}.

\bibitem{PARIKH2024112622}
\bibinfo{author}{Parikh, S.~J.} \emph{et~al.}
\newblock \bibinfo{title}{Structural and biophysical analysis of cytochrome p450 2c9*14 and *27 variants in complex with losartan}.
\newblock \emph{\bibinfo{journal}{Journal of Inorganic Biochemistry}} \textbf{\bibinfo{volume}{258}}, \bibinfo{pages}{112622} (\bibinfo{year}{2024}).
\newblock \urlprefix\url{https://www.sciencedirect.com/science/article/pii/S0162013424001466}.

\bibitem{Parikh2024Structural}
\bibinfo{author}{Parikh, S.~J.} \emph{et~al.}
\newblock \bibinfo{title}{Structural and biophysical analysis of cytochrome p450 2c9*14 and *27 variants in complex with losartan}.
\newblock \emph{\bibinfo{journal}{Journal of inorganic biochemistry}} \textbf{\bibinfo{volume}{258}}, \bibinfo{pages}{112622} (\bibinfo{year}{2024}).
\newblock \bibinfo{note}{[Online; accessed 2024-12-17]}.

\bibitem{Rao2020.12.15.422761}
\bibinfo{author}{Rao, R.}, \bibinfo{author}{Meier, J.}, \bibinfo{author}{Sercu, T.}, \bibinfo{author}{Ovchinnikov, S.} \& \bibinfo{author}{Rives, A.}
\newblock \bibinfo{title}{Transformer protein language models are unsupervised structure learners}.
\newblock \emph{\bibinfo{journal}{bioRxiv}}  (\bibinfo{year}{2020}).
\newblock \urlprefix\url{https://www.biorxiv.org/content/early/2020/12/15/2020.12.15.422761}.

\bibitem{sieg2019need}
\bibinfo{author}{Sieg, J.}, \bibinfo{author}{Flachsenberg, F.} \& \bibinfo{author}{Rarey, M.}
\newblock \bibinfo{title}{In need of bias control: evaluating chemical data for machine learning in structure-based virtual screening}.
\newblock \emph{\bibinfo{journal}{Journal of chemical information and modeling}} \textbf{\bibinfo{volume}{59}}, \bibinfo{pages}{947--961} (\bibinfo{year}{2019}).

\bibitem{gorantla_proteins_2024}
\bibinfo{author}{Gorantla, R.}, \bibinfo{author}{Kubincová, A.}, \bibinfo{author}{Weiße, A.~Y.} \& \bibinfo{author}{Mey, A. S. J.~S.}
\newblock \bibinfo{title}{From {Proteins} to {Ligands}: {Decoding} {Deep} {Learning} {Methods} for {Binding} {Affinity} {Prediction}}.
\newblock \emph{\bibinfo{journal}{Journal of Chemical Information and Modeling}} \textbf{\bibinfo{volume}{64}}, \bibinfo{pages}{2496--2507} (\bibinfo{year}{2024}).
\newblock \urlprefix\url{https://doi.org/10.1021/acs.jcim.3c01208}.
\newblock \bibinfo{note}{Publisher: American Chemical Society}.

\bibitem{hauser2017trends}
\bibinfo{author}{Hauser, A.~S.}, \bibinfo{author}{Attwood, M.~M.}, \bibinfo{author}{Rask-Andersen, M.}, \bibinfo{author}{Schi{\"o}th, H.~B.} \& \bibinfo{author}{Gloriam, D.~E.}
\newblock \bibinfo{title}{Trends in gpcr drug discovery: new agents, targets and indications}.
\newblock \emph{\bibinfo{journal}{Nature reviews Drug discovery}} \textbf{\bibinfo{volume}{16}}, \bibinfo{pages}{829--842} (\bibinfo{year}{2017}).

\bibitem{qiu_systematic_2024}
\bibinfo{author}{Qiu, Y.} \emph{et~al.}
\newblock \bibinfo{title}{Systematic characterization of multi-omics landscape between gut microbial metabolites and {GPCRome} in {Alzheimer}’s disease}.
\newblock \emph{\bibinfo{journal}{Cell Reports}} \textbf{\bibinfo{volume}{43}} (\bibinfo{year}{2024}).
\newblock \urlprefix\url{https://doi.org/10.1016/j.celrep.2024.114128}.
\newblock \bibinfo{note}{Publisher: Elsevier}.

\bibitem{Jansen2019Genome}
\bibinfo{author}{Jansen, I.~E.} \emph{et~al.}
\newblock \bibinfo{title}{Genome-wide meta-analysis identifies new loci and functional pathways influencing alzheimer’s disease risk}.
\newblock \emph{\bibinfo{journal}{Nature Genetics}} \textbf{\bibinfo{volume}{51}}, \bibinfo{pages}{404--413} (\bibinfo{year}{2019}).
\newblock \urlprefix\url{http://dx.doi.org/10.1038/s41588-018-0311-9}.

\bibitem{Wightman2021genome}
\bibinfo{author}{Wightman, D.~P.} \emph{et~al.}
\newblock \bibinfo{title}{A genome-wide association study with 1,126,563 individuals identifies new risk loci for alzheimer's disease}.
\newblock \emph{\bibinfo{journal}{Nature genetics}} \textbf{\bibinfo{volume}{53}}, \bibinfo{pages}{1276--1282} (\bibinfo{year}{2021}).
\newblock \bibinfo{note}{[Online; accessed 2024-10-17]}.

\bibitem{Zhou2021AlzGPS}
\bibinfo{author}{Zhou, Y.} \emph{et~al.}
\newblock \bibinfo{title}{Alzgps: a genome-wide positioning systems platform to catalyze multi-omics for alzheimer’s drug discovery}.
\newblock \emph{\bibinfo{journal}{Alzheimer's Research \& Therapy}} \textbf{\bibinfo{volume}{13}} (\bibinfo{year}{2021}).
\newblock \urlprefix\url{http://dx.doi.org/10.1186/s13195-020-00760-w}.

\bibitem{Han2021metabolomics}
\bibinfo{author}{Han, S.} \emph{et~al.}
\newblock \bibinfo{title}{A metabolomics pipeline for the mechanistic interrogation of the gut microbiome}.
\newblock \emph{\bibinfo{journal}{Nature}} \textbf{\bibinfo{volume}{595}}, \bibinfo{pages}{415--420} (\bibinfo{year}{2021}).
\newblock \urlprefix\url{http://dx.doi.org/10.1038/s41586-021-03707-9}.

\bibitem{Coletto2022role}
\bibinfo{author}{Coletto, E.} \emph{et~al.}
\newblock \bibinfo{title}{The role of the mucin-glycan foraging \textit{Ruminococcus gnavus} in the communication between the gut and the brain}.
\newblock \emph{\bibinfo{journal}{Gut microbes}} \textbf{\bibinfo{volume}{14}}, \bibinfo{pages}{2073784} (\bibinfo{year}{2022}).
\newblock \bibinfo{note}{[Online; accessed 2024-10-17]}.

\bibitem{dos2020Impact}
\bibinfo{author}{dos Santos~Guilherme, M.} \emph{et~al.}
\newblock \bibinfo{title}{Impact of acute and chronic amyloid-{$\beta$} peptide exposure on gut microbial commensals in the mouse}.
\newblock \emph{\bibinfo{journal}{Frontiers in Microbiology}} \textbf{\bibinfo{volume}{11}} (\bibinfo{year}{2020}).
\newblock \urlprefix\url{http://dx.doi.org/10.3389/fmicb.2020.01008}.

\bibitem{Averill2023antioxidant}
\bibinfo{author}{Averill-Bates, D.~A.}
\newblock \bibinfo{title}{The antioxidant glutathione}.
\newblock \emph{\bibinfo{journal}{Vitamins and hormones}} \textbf{\bibinfo{volume}{121}}, \bibinfo{pages}{109--141} (\bibinfo{year}{2023}).
\newblock \bibinfo{note}{[Online; accessed 2024-10-17]}.

\bibitem{Gould2019Impact}
\bibinfo{author}{Gould, R.~L.} \& \bibinfo{author}{Pazdro, R.}
\newblock \bibinfo{title}{Impact of supplementary amino acids, micronutrients, and overall diet on glutathione homeostasis}.
\newblock \emph{\bibinfo{journal}{Nutrients}} \textbf{\bibinfo{volume}{11}}, \bibinfo{pages}{1056} (\bibinfo{year}{2019}).
\newblock \bibinfo{note}{[Online; accessed 2024-10-17]}.

\bibitem{Mandal2019Cognitive}
\bibinfo{author}{Mandal, P.~K.}, \bibinfo{author}{Shukla, D.}, \bibinfo{author}{Tripathi, M.} \& \bibinfo{author}{Ersland, L.}
\newblock \bibinfo{title}{Cognitive improvement with glutathione supplement in alzheimer's disease: A way forward}.
\newblock \emph{\bibinfo{journal}{Journal of Alzheimer's disease : JAD}} \textbf{\bibinfo{volume}{68}}, \bibinfo{pages}{531--535} (\bibinfo{year}{2019}).
\newblock \bibinfo{note}{[Online; accessed 2024-10-17]}.

\bibitem{Minhas2024Restoring}
\bibinfo{author}{Minhas, P.~S.} \emph{et~al.}
\newblock \bibinfo{title}{Restoring hippocampal glucose metabolism rescues cognition across alzheimer's disease pathologies}.
\newblock \emph{\bibinfo{journal}{Science (New York, N.Y.)}} \textbf{\bibinfo{volume}{385}}, \bibinfo{pages}{eabm6131} (\bibinfo{year}{2024}).
\newblock \bibinfo{note}{[Online; accessed 2024-10-17]}.

\bibitem{Jung2022Gut}
\bibinfo{author}{Jung, J.~H.} \emph{et~al.}
\newblock \bibinfo{title}{Gut microbiome alterations in preclinical alzheimer's disease}.
\newblock \emph{\bibinfo{journal}{PloS one}} \textbf{\bibinfo{volume}{17}}, \bibinfo{pages}{e0278276} (\bibinfo{year}{2022}).
\newblock \bibinfo{note}{[Online; accessed 2024-10-17]}.

\bibitem{chen2019hidden}
\bibinfo{author}{Chen, L.} \emph{et~al.}
\newblock \bibinfo{title}{Hidden bias in the dud-e dataset leads to misleading performance of deep learning in structure-based virtual screening}.
\newblock \emph{\bibinfo{journal}{PloS one}} \textbf{\bibinfo{volume}{14}}, \bibinfo{pages}{e0220113} (\bibinfo{year}{2019}).

\bibitem{wang2020makes}
\bibinfo{author}{Wang, W.}, \bibinfo{author}{Tran, D.} \& \bibinfo{author}{Feiszli, M.}
\newblock \bibinfo{editor}{Boult, T.}, \bibinfo{editor}{Medioni, G.} \& \bibinfo{editor}{Zabih, R.} (eds) \emph{\bibinfo{title}{What makes training multi-modal classification networks hard?}}
\newblock (eds \bibinfo{editor}{Boult, T.}, \bibinfo{editor}{Medioni, G.} \& \bibinfo{editor}{Zabih, R.}) \emph{\bibinfo{booktitle}{2020 IEEE/CVF Conference on Computer Vision and Pattern Recognition (CVPR)}}, \bibinfo{pages}{12692--12702} (\bibinfo{publisher}{IEEE}, \bibinfo{year}{2020}).

\bibitem{wu2022characterizing}
\bibinfo{author}{Wu, N.}, \bibinfo{author}{Jastrzebski, S.}, \bibinfo{author}{Cho, K.} \& \bibinfo{author}{Geras, K.~J.}
\newblock \bibinfo{editor}{Chaudhuri, K.} \emph{et~al.} (eds) \emph{\bibinfo{title}{Characterizing and overcoming the greedy nature of learning in multi-modal deep neural networks}}.
\newblock (eds \bibinfo{editor}{Chaudhuri, K.} \emph{et~al.}) \emph{\bibinfo{booktitle}{Proceedings of the 39th International Conference on Machine Learning}}, Vol. \bibinfo{volume}{162} of \emph{\bibinfo{series}{Proceedings of Machine Learning Research}}, \bibinfo{pages}{24043--24055} (\bibinfo{publisher}{PMLR}, \bibinfo{year}{2022}).
\newblock \urlprefix\url{https://proceedings.mlr.press/v162/wu22d.html}.

\bibitem{zhang2024multimodal}
\bibinfo{author}{Zhang, Q.} \emph{et~al.}
\newblock \bibinfo{title}{Multimodal fusion on low-quality data: A comprehensive survey}.
\newblock \emph{\bibinfo{journal}{arXiv preprint arXiv:2404.18947}}  (\bibinfo{year}{2024}).

\bibitem{zhanggoltsgarcia2024}
\bibinfo{author}{Zhang, L.}, \bibinfo{author}{Golts, A.} \& \bibinfo{author}{Lopez~Garcia, V.}
\newblock \bibinfo{title}{Molecular representations for drug discovery} (\bibinfo{year}{2024}).
\newblock \bibinfo{note}{Accepted}.

\bibitem{zhou2023unimol}
\bibinfo{author}{Zhou, G.} \emph{et~al.}
\newblock \bibinfo{title}{Uni-mol: A universal 3d molecular representation learning framework}  (\bibinfo{year}{2023}).
\newblock \urlprefix\url{https://openreview.net/forum?id=6K2RM6wVqKu}.

\bibitem{cheng2024molecular}
\bibinfo{author}{Cheng, F.} \emph{et~al.}
\newblock \bibinfo{title}{A molecular video-derived foundation model streamlines scientific drug discovery}.
\newblock \emph{\bibinfo{journal}{Research Square}}  (\bibinfo{year}{2024}).

\bibitem{kinasedocking2024}
\bibinfo{author}{Backenköhler, M.}, \bibinfo{author}{Groß, J.}, \bibinfo{author}{Wolf, V.} \& \bibinfo{author}{Volkamer, A.}
\newblock \bibinfo{title}{Guided docking as a data generation approach facilitates structure-based machine learning on kinases}.
\newblock \emph{\bibinfo{journal}{Journal of Chemical Information and Modeling}} \textbf{\bibinfo{volume}{64}}, \bibinfo{pages}{4009--4020} (\bibinfo{year}{2024}).
\newblock \urlprefix\url{https://doi.org/10.1021/acs.jcim.4c00055}.
\newblock \bibinfo{note}{PMID: 38751014}.

\bibitem{kaplan2020scaling}
\bibinfo{author}{Kaplan, J.} \emph{et~al.}
\newblock \bibinfo{title}{Scaling laws for neural language models}.
\newblock \emph{\bibinfo{journal}{arXiv preprint arXiv:2001.08361}}  (\bibinfo{year}{2020}).

\bibitem{hu_strategies_2020}
\bibinfo{author}{Hu, W.} \emph{et~al.}
\newblock \bibinfo{title}{Strategies for {Pre}-training {Graph} {Neural} {Networks}} (\bibinfo{year}{2020}).
\newblock \urlprefix\url{http://arxiv.org/abs/1905.12265}.
\newblock \bibinfo{note}{ArXiv:1905.12265 [cs, stat]}.

\bibitem{weininger1988smiles}
\bibinfo{author}{Weininger, D.}
\newblock \bibinfo{title}{Smiles, a chemical language and information system. 1. introduction to methodology and encoding rules}.
\newblock \emph{\bibinfo{journal}{Journal of chemical information and computer sciences}} \textbf{\bibinfo{volume}{28}}, \bibinfo{pages}{31--36} (\bibinfo{year}{1988}).

\bibitem{schwaller2018found}
\bibinfo{author}{Schwaller, P.}, \bibinfo{author}{Gaudin, T.}, \bibinfo{author}{Lanyi, D.}, \bibinfo{author}{Bekas, C.} \& \bibinfo{author}{Laino, T.}
\newblock \bibinfo{title}{“found in translation”: predicting outcomes of complex organic chemistry reactions using neural sequence-to-sequence models}.
\newblock \emph{\bibinfo{journal}{Chemical science}} \textbf{\bibinfo{volume}{9}}, \bibinfo{pages}{6091--6098} (\bibinfo{year}{2018}).

\bibitem{he2016deep}
\bibinfo{author}{He, K.}, \bibinfo{author}{Zhang, X.}, \bibinfo{author}{Ren, S.} \& \bibinfo{author}{Sun, J.}
\newblock \bibinfo{title}{Deep residual learning for image recognition} \bibinfo{pages}{770--778} (\bibinfo{year}{2016}).

\bibitem{su2024roformer}
\bibinfo{author}{Su, J.} \emph{et~al.}
\newblock \bibinfo{title}{Roformer: Enhanced transformer with rotary position embedding}.
\newblock \emph{\bibinfo{journal}{Neurocomputing}} \textbf{\bibinfo{volume}{568}}, \bibinfo{pages}{127063} (\bibinfo{year}{2024}).

\bibitem{katharopoulos_et_al_2020}
\bibinfo{author}{Katharopoulos, A.}, \bibinfo{author}{Vyas, A.}, \bibinfo{author}{Pappas, N.} \& \bibinfo{author}{Fleuret, F.}
\newblock \bibinfo{title}{Transformers are rnns: Fast autoregressive transformers with linear attention}  (\bibinfo{year}{2020}).

\bibitem{bemis_properties_1996}
\bibinfo{author}{Bemis, G.~W.} \& \bibinfo{author}{Murcko, M.~A.}
\newblock \bibinfo{title}{The {Properties} of {Known} {Drugs}. 1. {Molecular} {Frameworks}}.
\newblock \emph{\bibinfo{journal}{Journal of Medicinal Chemistry}} \textbf{\bibinfo{volume}{39}}, \bibinfo{pages}{2887--2893} (\bibinfo{year}{1996}).
\newblock \urlprefix\url{https://doi.org/10.1021/jm9602928}.
\newblock \bibinfo{note}{Publisher: American Chemical Society}.

\bibitem{liaw2018tune}
\bibinfo{author}{Liaw, R.} \emph{et~al.}
\newblock \bibinfo{title}{Tune: A research platform for distributed model selection and training}.
\newblock \emph{\bibinfo{journal}{arXiv preprint arXiv:1807.05118}}  (\bibinfo{year}{2018}).

\bibitem{Jumper2021Highly}
\bibinfo{author}{Jumper, J.} \emph{et~al.}
\newblock \bibinfo{title}{Highly accurate protein structure prediction with alphafold}.
\newblock \emph{\bibinfo{journal}{Nature}} \textbf{\bibinfo{volume}{596}}, \bibinfo{pages}{583--589} (\bibinfo{year}{2021}).
\newblock \urlprefix\url{http://dx.doi.org/10.1038/s41586-021-03819-2}.

\bibitem{Eberhardt2021AutoDock}
\bibinfo{author}{Eberhardt, J.}, \bibinfo{author}{Santos-Martins, D.}, \bibinfo{author}{Tillack, A.~F.} \& \bibinfo{author}{Forli, S.}
\newblock \bibinfo{title}{Autodock vina 1.2.0: New docking methods, expanded force field, and python bindings}.
\newblock \emph{\bibinfo{journal}{Journal of Chemical Information and Modeling}} \textbf{\bibinfo{volume}{61}}, \bibinfo{pages}{3891--3898} (\bibinfo{year}{2021}).

\end{thebibliography}

\end{document}


\title[Article Title]{Supplementary Tables and Figures for `Multi-view biomedical foundation models for molecule-target and property prediction'}


\author[1]{\fnm{Parthasarathy} \sur{Suryanarayanan}}
\equalcont{These authors contributed equally to this work.}

\author[2,3]{\fnm{Yunguang} \sur{Qiu}}
\equalcont{These authors contributed equally to this work.}

\author[4]{\fnm{Shreyans} \sur{Sethi}}

\author[1]{\fnm{Diwakar} \sur{Mahajan}}

\author[1]{\fnm{Hongyang} \sur{Li}}

\author[2]{\fnm{Yuxin} \sur{Yang}}

\author[1]{\fnm{Elif} \sur{Eyigoz}}

\author[1]{\fnm{Aldo} \sur{Guzm\'an S\'aenz}}

\author[1]{\fnm{Daniel E.} \sur{Platt}}

\author[1]{\fnm{Timothy H.} \sur{Rumbell}}

\author[5]{\fnm{Kenney} \sur{Ng}}

\author[1]{\fnm{Sanjoy} \sur{Dey}}

\author[1]{\fnm{Myson} \sur{Burch}}

\author[5]{\fnm{Bum Chul} \sur{Kwon}}

\author[1]{\fnm{Pablo} \sur{Meyer}}

\author[2,3,6]{\fnm{Feixiong} \sur{Cheng}}

\author[1]{\fnm{Jianying} \sur{Hu}}

\author*[1]{\fnm{Joseph A.} \sur{Morrone}} \email{jamorron@us.ibm.com}

\affil[1]{\orgname{IBM TJ Watson Research Center}, \orgaddress{\street{1101 Kitchawan Rd}, \city{Yorktown Heights},  \state{NY}, \postcode{10598}, \country{USA}}}

\affil[2]{\orgdiv{Cleveland Clinic Genome Center}, \orgname{Lerner Research Institute, Cleveland Clinic}, \orgaddress{ \city{Cleveland},  \state{OH}, \postcode{44195}, \country{USA}}}

\affil[3]{\orgdiv{Genomic Medicine Institute}, \orgname{Lerner Research Institute, Cleveland Clinic}, \orgaddress{ \city{Cleveland}, \state{OH},  \postcode{44195}, \country{USA}}}

\affil[4]{\orgname{IBM Research - Almaden}, \orgaddress{\street{650 Harry Rd}, \city{San Jose},  \state{CA}, \postcode{95120}, \country{USA}}}

\affil[5]{\orgname{IBM Research}, \orgaddress{\street{314 Main St}}, \city{Cambridge},  \state{MA}, \postcode{02142}, \country{USA}}

\affil[6]{\orgdiv{Department of Molecular Medicine}, \orgname{Cleveland Clinic Lerner College of Medicine, Case Western Reserve University}, \city{Cleveland}, \state{OH}, \postcode{44195}, \country{USA}}

\maketitle
\newpage

\begin{figure}
    \centering
    \includegraphics[width=1.0\linewidth]{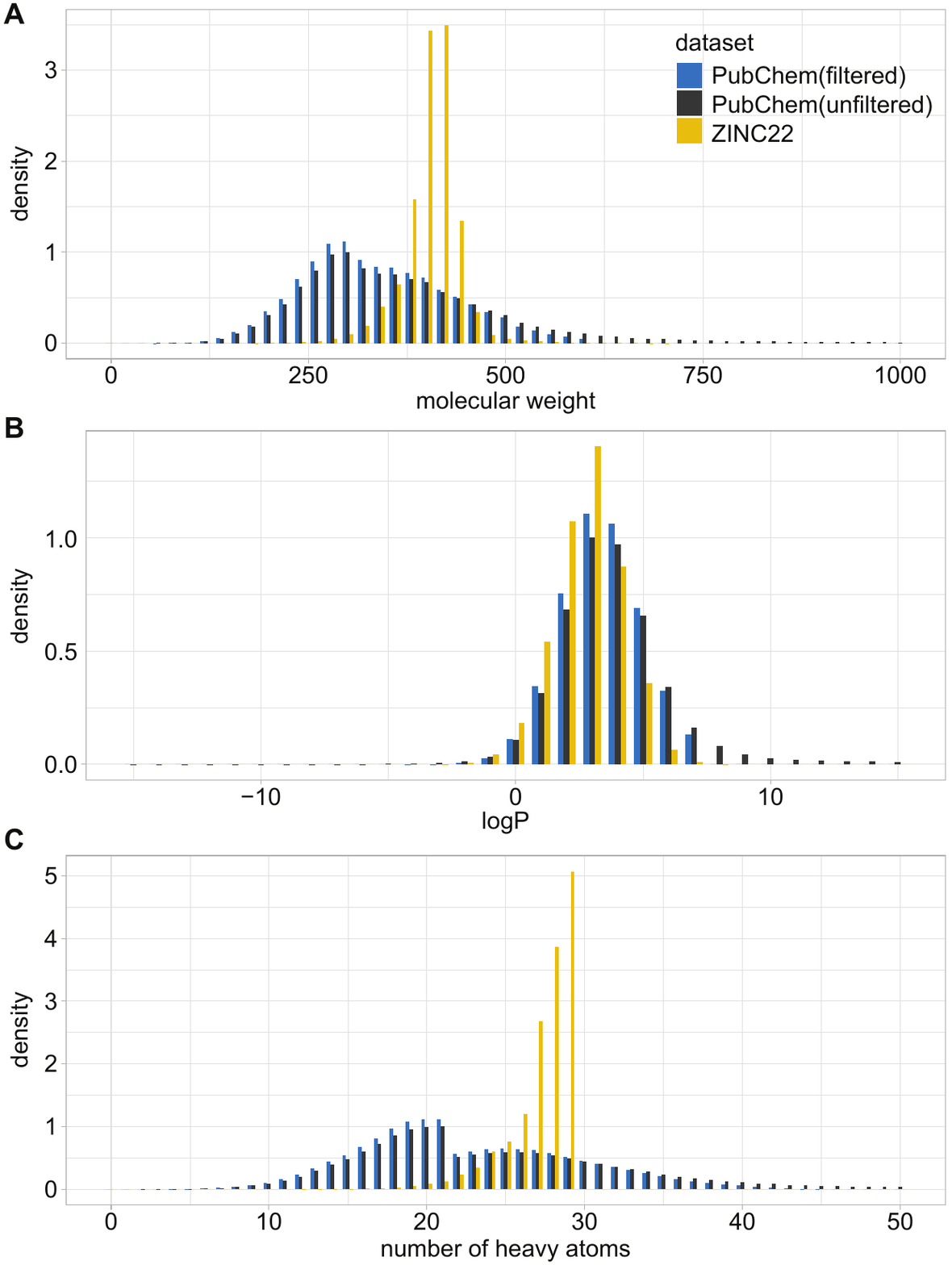}
    \caption{Molecular properties of the pre-training dataset.  Histograms of molecular weight (A), logP (B), and number of heavy atoms (C) are shown.  Distributions are plotted for the filtered PubChem set used to build our foundation models (light blue), the full PubChem set (dark blue) and the molecules of ZINC22 that comprise the remainder of our dataset (yellow).}
    \label{fig:pretrain_hist_1}
\end{figure}

\begin{figure}
    \centering
    \includegraphics[width=1.0\linewidth]{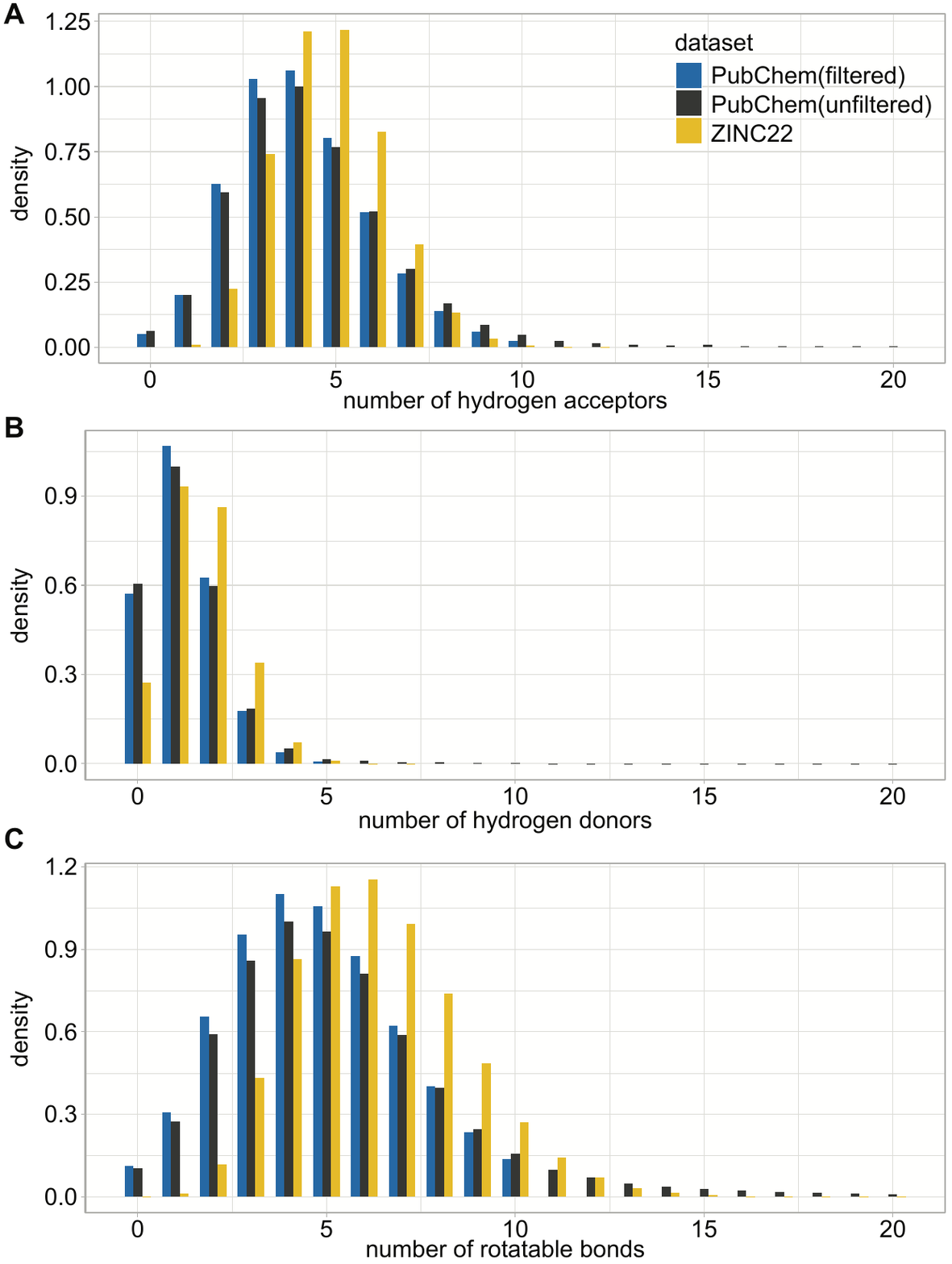}
    \caption{Molecular properties of the pre-training dataset.  Number of hydrogen bond acceptors (A), number of hydrogen bond donors  (B), and number of rotatable bonds (C) are shown.  Distributions are plotted for the filtered PubChem set used to build our foundation models (light blue), the full PubChem set (dark blue) and the molecules of ZINC22 that comprise the remainder of our dataset (yellow).}
    \label{fig:pretrain_hist_2}
\end{figure}

\begin{figure}
    \centering
    \includegraphics[width=1.0\linewidth]{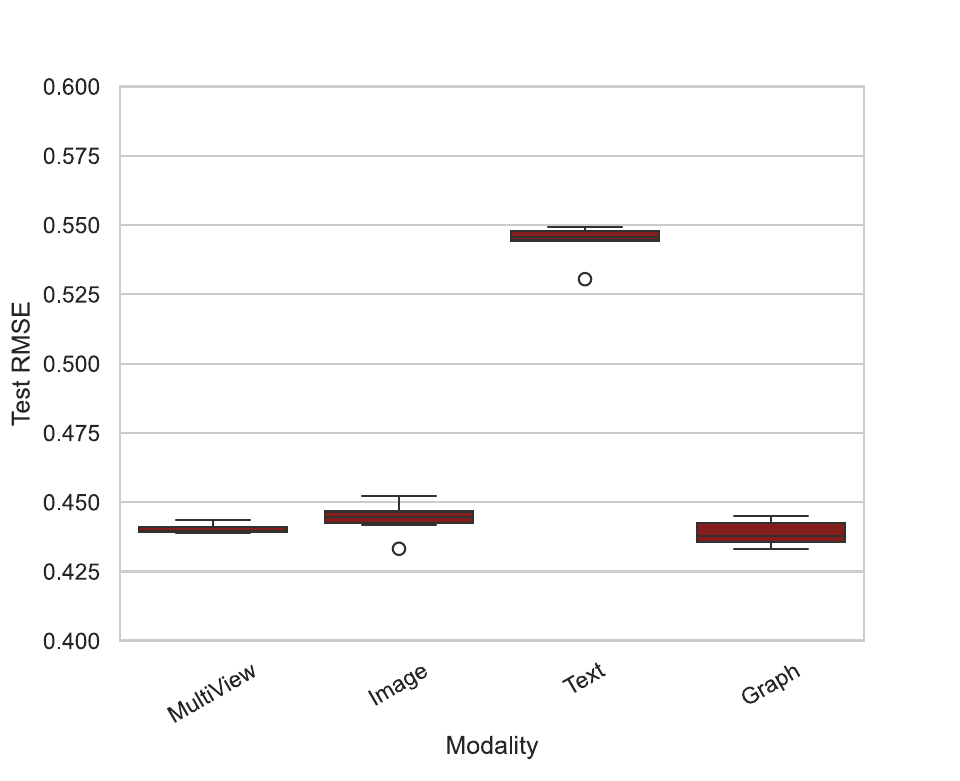}
    \caption{RMSE for multi-view, Image, Graph, Text model fine-tuned on the Davis dataset.  The Davis dataset contains the interaction of 72 kinase inhibitors with 442 kinases.  Results are compared with reference value of Gorantla, et al. which is $\approx 0.48$}
    \label{fig:davis}
\end{figure}

\begin{figure}
    \centering
    \includegraphics[width=1.0\linewidth]{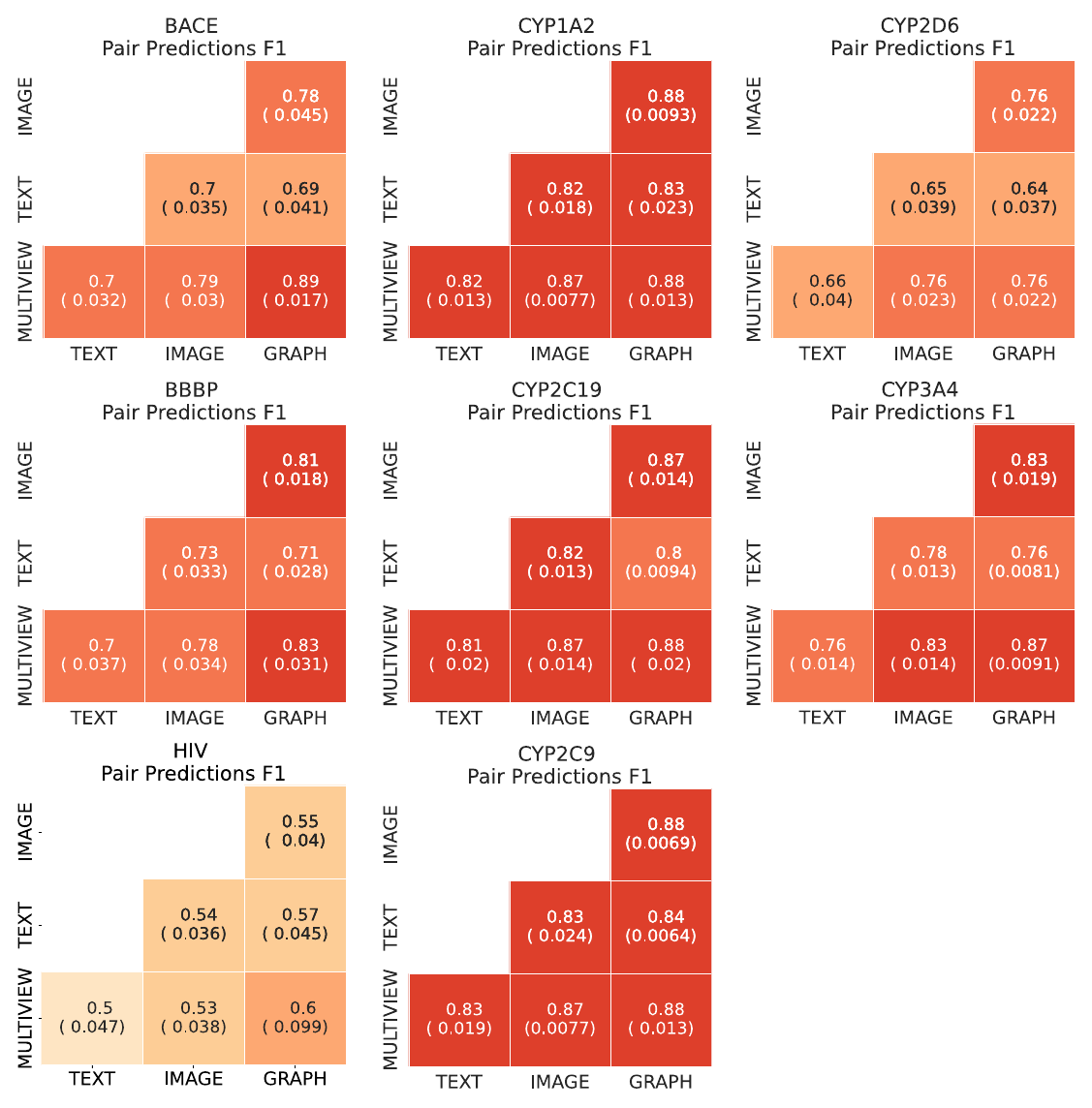}
    \caption{$F1$ scores are shown between pairs of single-view and multi-view models. Results are shown for single task classification datasets.  Higher F1 values indicate more agreement between the models vs. disagreements. In the case of CYP models the agreement between multi-view - Image, multi-view Graph and Image-Graph is nearly the same, indicating that the models yield largely the same outcomes.  For other models, the Graph - multi-view yields the highest $F1$ score as expected from Fig. 2C in the main text.}
    \label{fig:stats_class}
\end{figure}

\begin{figure}
    \centering
    \includegraphics[width=1.0\linewidth]{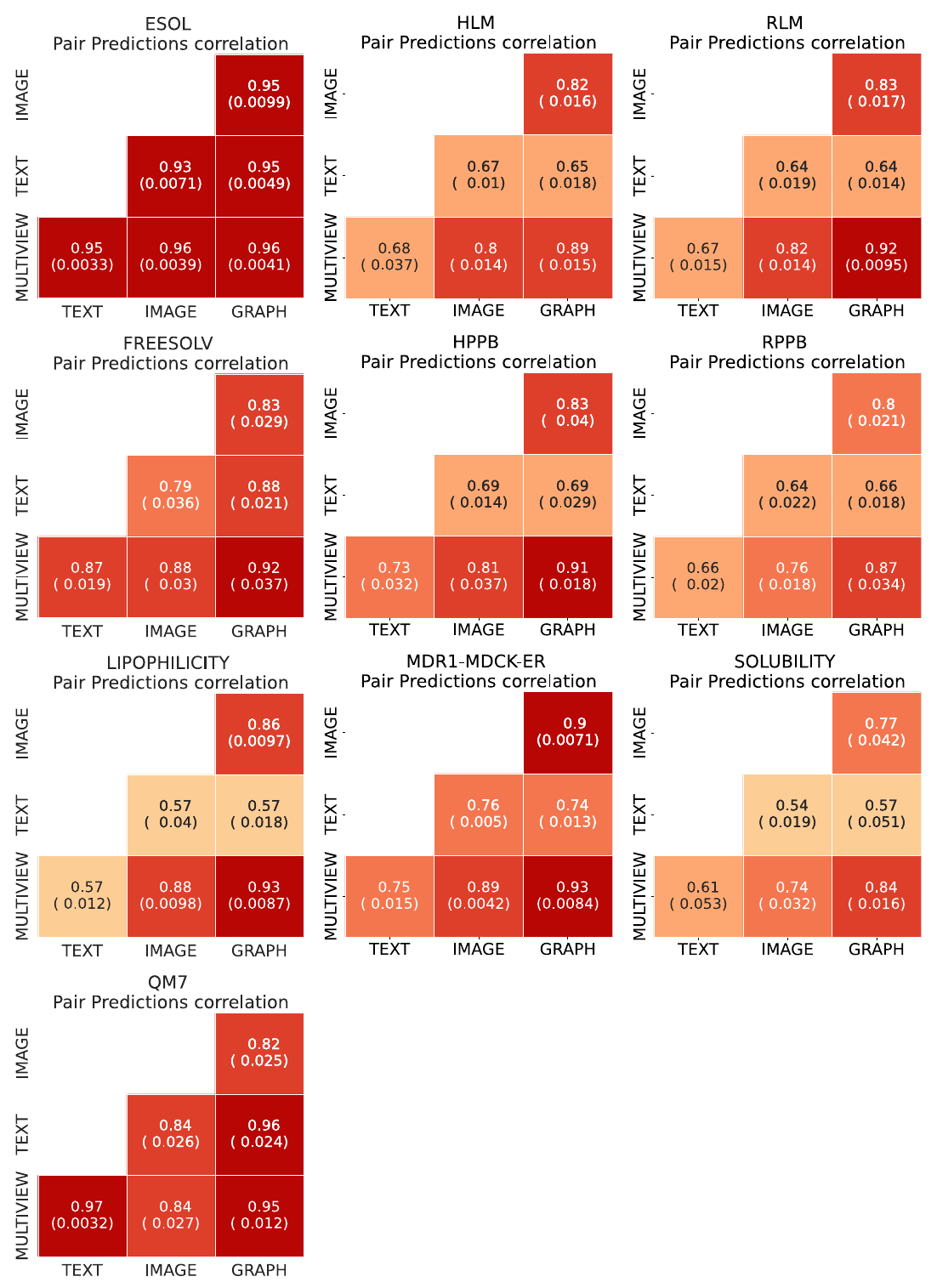}
    \caption{Pearson corelation scores are shown between pairs of models for regression tasks.  A higher number indicates greater corelation in the model predictions. For ComputationalADME tasks, Graph is the most correlated to multi-view as expected from Fig. 2C in the main text.  In the case of QM7 the correlation between Text and multi-view is relatively high compared with other datasets.}
    \label{fig:stats_regr}
\end{figure}

\begin{figure}
    \centering
    \includegraphics[width=1\linewidth]{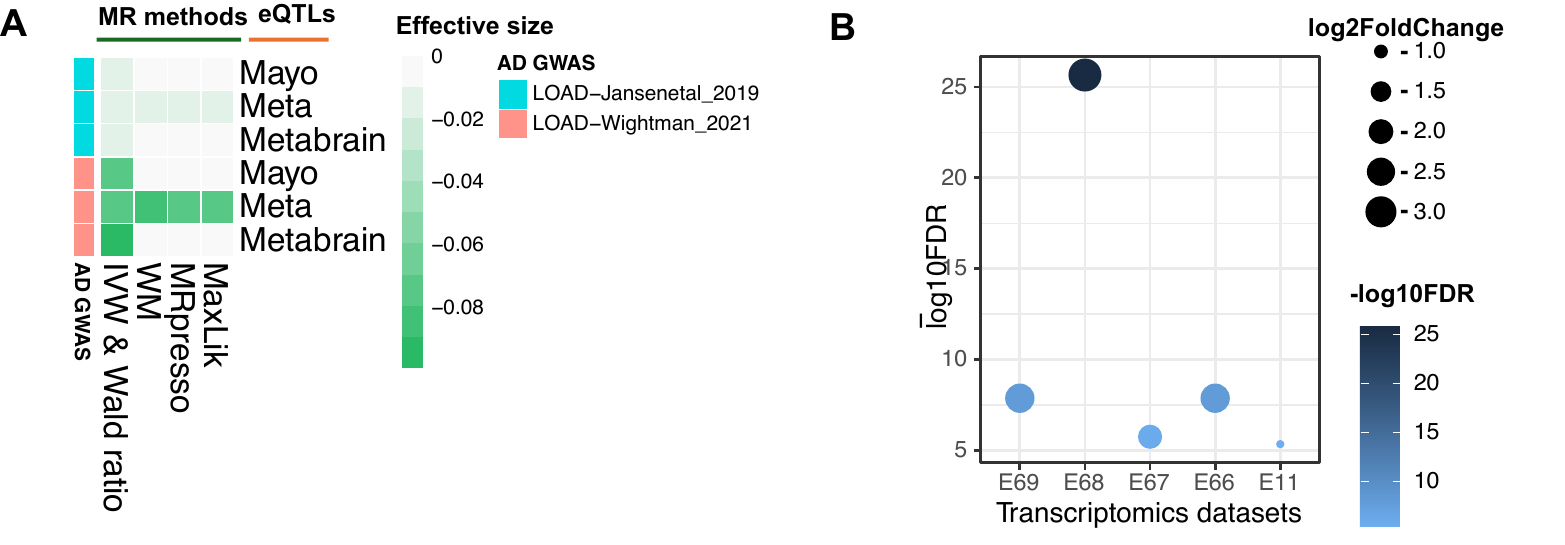}
    \caption{Genetics or multi-omics evidenced AD-related GPCRs. (A) Genetics-informed AD-related FPR1 by Mendelian Randomization (MR). In total, three genomic datasets (Mayo, Meta and Metabrain) in human cortex region were used. effective size $< 0$ indicates elevated expression level of GPCRs decrease the likelihood of AD. Only data with FDR $< 0.05$ was shown. (B) Transcriptomics-informed AD-related ADA2A by investigating AlzGPS database (https://alzgps.lerner.ccf.org/).}
    \label{fig:ad_analysis}
\end{figure}

\begin{figure}
    \centering
    \includegraphics[width=1\linewidth]{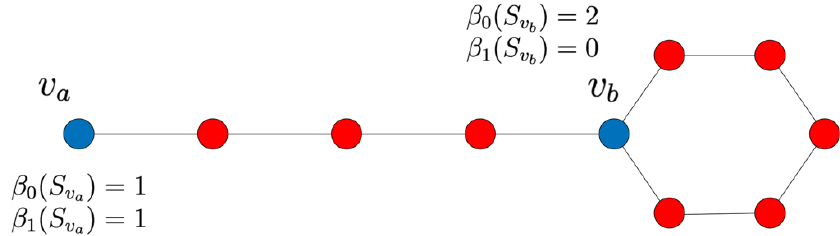}
    \caption{Graph topology depiction that illustrates Betti numbers at a given node.  Refer to Sec. 4.4 for more details.}
    \label{fig:topology}
\end{figure}

\FloatBarrier


\begin{table}
    \centering
    \resizebox{\textwidth}{!}{\begin{tabular}{cccccccc}
        \toprule
        Task & Group & Description	 & $N_\text{samples}$	& $N_\text{tasks}$	& type & metric & split \\
        \midrule
        BACE & MoleculeNet & Inhibition of human beta secretase 1	 & 1513	& 1	& C & roc-auc & size-ordered scaffold \\
        BBBP & MoleculeNet & Blood brain barrier penetration	& 2039 & 1 & C & roc-auc & size-ordered scaffold \\
        HIV & MoleculeNet & Inhibition of HIV viral replication & 41127 & 1 & C & roc-auc& size-ordered scaffold  \\
        MUV &  MoleculeNet & PubChem activities & 93087 & 17 & C & roc-auc & size-ordered scaffold  \\
        TOX21 & MoleculeNet & Toxicity data & 7831	& 12 & C & roc-auc & size-ordered scaffold  \\
        TOXCAST & MoleculeNet & Toxicity data & 8576 & 617	& C & roc-auc & size-ordered scaffold  \\
        ESOL & MoleculeNet & Water solubility data for organics &	1128 &	1 &	R  & rmse & size-ordered scaffold \\
        FREESOLV & MoleculeNet & Hydration free energy & 642 & 1 & R & rmse & size-ordered scaffold  \\
        LIPOPHILICITY & MoleculeNet & Octanol/water distribution coeff. & 4200 & 1 & R & rmse & size-ordered scaffold  \\
        QM7 & MoleculeNet & Electronic properties from DFT & 6830 & 1 & R & mae & size-ordered scaffold  \\
        CYP1A2 & CYP & Inhibition of CYP1A2 isoform & 11725 & 1 & C & roc-auc & balanced scaffold \\
        CYP2C19 & CYP & Inhibition of CYP2C19 isoform & 11533 & 1 & C & roc-auc & balanced scaffold \\
        CYP2C9 & CYP & Inhibition of CYP2C9 isoform & 11722 & 1 & C & roc-auc & balanced scaffold \\
        CYP2D6 & CYP & Inhibition of CYP2D6 isoform & 11540 & 1 & C & roc-auc & balanced scaffold \\
        CYP3A4 & CYP & Inhibition of CYP3A4 isoform & 11118 & 1 & C & roc-auc & balanced scaffold \\
        HLM & ComputationalADME & Human liver microsomal stability & 3087 & 1 & R & rmse & random \\
        HPPB & ComputationalADME & Human plasma protein binding & 1801 & 1 & R & rmse & random \\
        MDR1-MDCK-ER & ComputationalADME & MDR1-MDCK efflux ratio & 2642 & 1 & R & rmse & random \\
        RLM & ComputationalADME & Rat liver microsomal stability & 3054 & 1 & R & rmse & random \\
        RPPB & ComputationalADME & Rat plasma protein binding & 879 & 1 & R & rmse & random \\
        SOLUBILITY & ComputationalADME & Water solubility data & 2173 & 1 & R & rmse & random \\
        \bottomrule
    \end{tabular}}
    \caption{Description of benchmark tasks that are plotted in Fig. 2 in the main text, including CYP inhibition, ComputationalADME, and selections from MoleculeNet. Task types are classification (C) or regression (R).}
    \label{tab:molnet_description}
\end{table}

\FloatBarrier
\newpage

\begin{table}
    \centering
    \begin{tabular}{lllll}
    \toprule
    dataset & ESOL & FREESOLV & LIPOPHILICITY & QM7 \\
    model &  &  &  &  \\
    \midrule
    GCN & 1.350(54) & 2.575(22) & 0.818(15) & 102.185(7768) \\
    Graph & 0.836(26) & 1.900(36) & 0.671(21) & 77.324(4371) \\
    Image & 0.934(2) & 2.426(322) & 0.701(8) & 95.584(333) \\
    Text & 0.889(22) & 2.422(9) & 1.024(32) & 73.858(3463) \\
    MultiView & 0.853(35) & 2.046(121) & 0.654(12) & 74.105(2883) \\
    \bottomrule
    \end{tabular}
    \caption{MoleculeNet Size-ordered scaffold results for MoleculNet regression tasks with $95\%$ confidence interval in parenthesis. Metric is  MAE for QM7, RMSE for other tasks.}
    \label{tab:molnet_scaffold_regr}
\end{table}

\begin{table}
    \centering
    \begin{tabular}{lllllll}
    \toprule
    dataset & BACE & BBBP & HIV & MUV & TOX21 & TOXCAST \\
    model &  &  &  &  &  &  \\
    \midrule
    GCN & 0.733(7) & 0.676(28) & 0.735(1) & 0.666(24) & 0.730(8) & 0.628(8) \\
    Graph & 0.795(21) & 0.656(11) & 0.787(14) & 0.810(12) & 0.770(8) & 0.683(4) \\
    Image & 0.766(25) & 0.684(1) & 0.764(23) & 0.647(25) & 0.754(7) & 0.659(5) \\
    Text & 0.709(27) & 0.681(7) & 0.743(15) & 0.673(8) & 0.744(7) & 0.639(2) \\
    MultiView & 0.798(23) & 0.706(19) & 0.778(11) & 0.794(14) & 0.770(3) & 0.664(8) \\
    \bottomrule
    \end{tabular}
    \caption{Computed ROC-AUC for test set for size-order scaffold split for selected MoleculeNet classification tasks with $95\%$ confidence interval in parenthesis.}
    \label{tab:molnet_scaffold_class}
\end{table}

\begin{table}
    \centering
    \begin{tabular}{llllll}
    \toprule
    dataset & CYP1A2 & CYP2C19 & CYP2C9 & CYP2D6 & CYP3A4 \\
    model &  &  &  &  &  \\
    \midrule
    GCN & 0.869(2) & 0.836(3) & 0.891(4) & 0.798(6) & 0.832(6) \\
    Graph & 0.910(2) & 0.867(4) & 0.914(6) & 0.827(7) & 0.891(2) \\
    Image & 0.896(3) & 0.862(4) & 0.896(3) & 0.824(2) & 0.865(6) \\
    Text & 0.855(2) & 0.807(3) & 0.877(2) & 0.759(6) & 0.804(7) \\
    MultiView & 0.899(2) & 0.864(5) & 0.900(7) & 0.820(3) & 0.888(5) \\
    \bottomrule
    \end{tabular}
    \caption{ROC-AUC for CYP inhibition tasks with $95\%$ confidence interval in parenthesis}
    \label{tab:cyp_table}
\end{table}

\begin{table}
    \centering
    \begin{tabular}{lllllll}
    \toprule
    dataset & HLM & HPPB & MDR1-MDCK-ER & RLM & RPPB & SOLUBILITY \\
    model &  &  &  &  &  &  \\
    \midrule
    GCN & 0.526(18) & 0.557(12) & 0.607(15) & 0.565(7) & 0.603(15) & 0.539(18) \\
    Graph & 0.421(9) & 0.486(5) & 0.464(7) & 0.492(6) & 0.560(1) & 0.482(12) \\
    Image & 0.441(6) & 0.545(34) & 0.465(6) & 0.524(14) & 0.616(11) & 0.538(9) \\
    Text & 0.526(4) & 0.611(7) & 0.558(4) & 0.599(5) & 0.630(7) & 0.563(9) \\
    MultiView & 0.435(1) & 0.490(13) & 0.488(9) & 0.495(9) & 0.551(18) & 0.508(1) \\
    \bottomrule
    \end{tabular}
    \caption{ComputationalADME random splitting results with $95\%$ confidence interval in parenthesis.  For these regression tasks  RMSE is the chosen metric.}
    \label{tab:adme_regr}
\end{table}

\FloatBarrier
\newpage

\begin{table}
    \centering
    \begin{tabular}{p{2.5in}p{3.2in}}
    \toprule
        Category & AD-related GPCRs \\
         \midrule
        Genetic evidence-supported &  FPR1 \\
         Strong Multi-omics evidence-supported \newline (at least differential expressed in 5 datasets) & ADA2A \\
         Weak multi-omics evidence-supported \newline (differential expressed in less than 5 datasets) & N: 31 \newline PE2R3, P2Y12, 5HT2A, HRH3, OPRK, EDNRA, ACM3, \newline CCR2, 5HT1D, NPY1R, 5HT1A, GASR, ACM2, ADA1D, \newline CNR1, PE2R4, MCHR1, ADRB1, ACM5, MTR1A, 5HT2C, \newline 5HT1B, DRD1, CCR5, 5HT1F, 5HT7R, DRD5, 5HT5A, \newline NPY2R, DRD2, NPY5R \\
         & \\
         \bottomrule
    \end{tabular}
    \caption{Genetics- or multi-omics-informed AD-related GPCRs}
    \label{tab:AD_targets}
\end{table}
\newpage

\FloatBarrier
\begin{table}
    \centering
    \begin{tabular}{lllll}
    \toprule
    dataset & ESOL & FREESOLV & LIPOPHILICITY & QM7 \\
    model &  &  &  &  \\
    \midrule
    MultiView & 0.853(35) & 2.046(121) & 0.654(12) &  74.1(29) \\
    Unprojected Gating & 0.961(38) & 2.192(51) & 0.694(16) & 83.8(59) \\
    Projected Gating & 0.939(14) & 2.251(62) & 0.693(9) & 73.8(24) \\
    Projected Gating with Feature & 0.953(17) & 2.225(6) & 0.687(14) & 72.7(50) \\
    \bottomrule
    \end{tabular}
    \caption{Regression metrics (RMSE/MAE) for default multi-view model compared against other late fusion approaches that were explored (see Sec. 4.5). }
    \label{tab:rmse-mae_multiview}
\end{table}

\begin{table}
    \centering
    \begin{tabular}{llllll}
    \toprule
    dataset & BACE & BBBP & HIV & MUV \\
    model &  &  &  &  &  \\
    \midrule
    MultiView & 0.798(23) & 0.706(19)  & 0.778(11) & 0.794(14) \\
    Unprojected Gating & 0.798(27) & 0.708(18) & 0.754(8) & 0.740(19) \\
    Projected Gating & 0.784(8) & 0.690(8) & 0.761(7) & 0.734(18) \\
    Projected Gating with Feature & 0.789(9) & 0.707(7) & 0.757(4) & 0.734(11) \\
    \bottomrule
    \end{tabular}
    \caption{Classification metrics (ROC-AUC) for default multi-view model compared against other late fusion approaches that were explored (see Sec. 4.5). }
    \label{tab:rocauc_multiview}
\end{table}